\documentclass[journal]{IEEEtran}
\usepackage{amsmath,amsfonts, amssymb}
\usepackage{algorithmic}
\usepackage{algorithm}
\usepackage{array}
\usepackage[caption=false,font=normalsize,labelfont=sf,textfont=sf]{subfig}
\usepackage{textcomp}
\usepackage{stfloats}
\usepackage{url}
\usepackage{verbatim}
\usepackage{graphicx}
\usepackage{cite}
\usepackage{bm}
\usepackage{pgfplots, pgfplotstable}
\usepackage{tikz}
\usepackage{soul}
\usepackage{booktabs}
\usepackage{multirow}

\pgfplotsset{compat=newest}
\usetikzlibrary{shapes, arrows, calc, patterns, matrix, positioning,fit,decorations.markings}
\usepackage{ifthen}
\pgfplotsset{plot coordinates/math parser=false}
\newlength\figureheight
\newlength\figurewidth 

\sethlcolor{yellow}

\newcommand{\etal}{\textit{et al.~}}

\newcommand{\Prob}{\mathrm{Pr}}

\newcommand{\QuantStep}{Q}
\newcommand{\loss}{\mathcal{L}}
\newcommand{\dist}{\mathcal{D}}
\newcommand{\distMSE}{\dist_\mathrm{MSE}}
\newcommand{\rate}{\mathcal{R}}
\newcommand{\lagrange}{\lambda}
\newcommand{\PSNR}{\mathrm{PSNR}}
\newcommand{\dB}{\,\ensuremath{\text{dB~}}}
\newcommand{\dBC}{\,\ensuremath{\text{dB}}}

\newcommand{\predMod}{\mathrm{p}}

\newcommand{\cur}{x}
\newcommand{\curV}{\bm{\cur}}
\newcommand{\pred}{\cur_\predMod}

\newcommand{\predV}{\bm{\cur}_\predMod}
\newcommand{\predBN}{\hat\cur_\predMod}
\newcommand{\predBNV}{\hat{\bm{\cur}}_\predMod}

\newcommand{\recV}{\tilde{\bm{\cur}}}
\newcommand{\res}{r}
\newcommand{\resV}{\bm{\res}}

\newcommand{\curD}{X}
\newcommand{\predD}{X_\mathrm{p}}
\newcommand{\resD}{R}
\newcommand{\recresD}{\tilde{R}}
\newcommand{\recD}{\tilde{X}}
\newcommand{\predBND}{\hat{X}_\mathrm{p}}

\newcommand{\RateRes}{\rate_\mathrm{Res}}
\newcommand{\RateCond}{\rate_\mathrm{Cond,ideal}}
\newcommand{\RateResCond}{\rate_\mathrm{CondRes,ideal}}
\newcommand{\RateCondBN}{\rate_\mathrm{Cond}}
\newcommand{\RateResCondBN}{\rate_\mathrm{CondRes}}
\newcommand{\PSetBN}{\mathcal{M}_\mathrm{BN}}
\newcommand{\PSet}{\mathcal{M}_0}

\newcommand{\latent}{y}

\newcommand{\intermedBefore}{c}

\newcommand{\recresV}{\tilde{\bm{\res}}}

\newcommand{\probMVE}{p}

\newcommand{\encoderNet}{f_e}
\newcommand{\decoderNet}{f_d}

\newcommand{\latentPredV}{\bm{y}_p}
\newcommand{\predChannels}{N_p}
\newcommand{\decChannels}{N_d}
\newcommand{\predPSNR}{\ensuremath{\PSNR_\mathrm{pred}}}
\newcommand{\BN}{B}
\newcommand{\funcEnc}{f_e}
\newcommand{\funcDec}{f_d}

\newcommand{\funcPredEnc}{f_p}

\newcommand{\hle}{\color{black}}

\newlength{\mylen}
\setbox1=\hbox{$\bullet$}\setbox2=\hbox{\tiny$\bullet$}
\begin{document}

\title{Conditional Residual Coding:\\ A Remedy for Bottleneck Problems in\\ Conditional Inter Frame Coding}

\author{\IEEEauthorblockN{Fabian Brand, \textit{Graduate Student Member,\,IEEE}, J\"urgen Seiler, \textit{Senior Member,\,IEEE},\\ and Andr\'e Kaup, \textit{Fellow,\,IEEE}}
	
	\vspace{0.1cm}
	\IEEEauthorblockA{Multimedia Communications and Signal Processing\\
		Friedrich-Alexander-Universit\"at Erlangen-N\"urnberg (FAU) \\
		Cauerstr. 7, 91058 Erlangen, Germany\vspace{-0.3cm}}}


\IEEEpubid{0000--0000/00\$00.00~\copyright~2021 IEEE}

\maketitle

\makeatletter
\def\ps@IEEEtitlepagestyle{%
	\def\@oddfoot{\mycopyrightnotice}%
	\def\@oddhead{\hbox{}\@IEEEheaderstyle\leftmark\hfil\thepage}\relax
	\def\@evenhead{\@IEEEheaderstyle\thepage\hfil\leftmark\hbox{}}\relax
	\def\@evenfoot{}%
}
\def\mycopyrightnotice{%
	\begin{minipage}{\textwidth}
		\scriptsize
		Copyright \copyright 2024 IEEE. Personal use of this material is permitted. However, permission to use this material for any other purposes must be obtained from the IEEE by sending an email to pubs-permissions@ieee.org.
	\end{minipage}
}
\makeatother

\begin{abstract}
Conditional coding is a new video coding paradigm enabled by neural-network-based compression. It can be shown that conditional coding is in theory better than the traditional residual coding, which is widely used in video compression standards like HEVC or VVC. However, on closer inspection, it becomes clear that conditional coders can suffer from information bottlenecks in the prediction path, i.e., that due to the data processing inequality not all information from the prediction signal can be passed to the reconstructed signal, thereby impairing the coder performance. In this paper we propose the conditional residual coding concept, which we derive from information theoretical properties of the conditional coder. This coder significantly reduces the influence of bottlenecks, while maintaining the theoretical performance of the conditional coder. We provide a theoretical analysis of the coding paradigm and demonstrate the performance of the conditional residual coder in a practical example. We show that conditional residual coders alleviate the disadvantages of conditional coders while being able to maintain their advantages over residual coders. In the spectrum of residual and conditional coding, we can therefore consider them as ``the best from both worlds''.
\end{abstract}

\begin{IEEEkeywords}
Video Compression, Conditional Coding, Conditional Autoencoder, Information Theory
\end{IEEEkeywords}

\section{Introduction}
For the last few years, neural-network-based image and video compression has drawn more and more attention in research. Early works in this field mainly include works on image compression~\cite{TodericiOH2016_VariableRateImage, BalleLS2017_Endendoptimized, BalleMS2018_Variationalimagecompression}, which was quickly adopted to video compression~\cite{LuOX2018_DVCEndend,HuLX2021_FVCNewFramework}. Most works on novel video compression techniques aim to replace the components known from traditional hybrid video coders, like HEVC~\cite{SullivanOH2012_OverviewHighEfficiency}, VVC~\cite{BrossWY2021_OverviewVersatileVideo}, or AV1~\cite{HanLM2021_TechnicalOverviewAV1}, with neural-network-based counterparts, maintaining the general paradigm of a combination of motion compensated prediction and residual coding. However, neural networks offer an exiting new possibility to change the paradigm from residual coding towards conditional coding, since neural networks can directly learn the statistical properties of the signals\cite{LadunePH2020_ModeNetModeSelection}.

\begin{figure}
	\centering
	\includegraphics{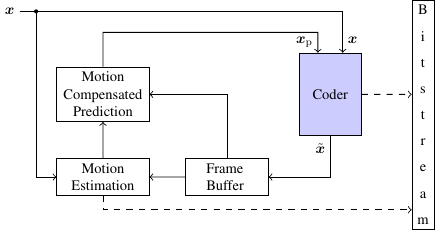}
	\caption{Schematic structure of an inter coder. The core element is the coder which compresses the frame $\curV$ with the help of a prediction signal $\predV$. The prediction signal was computed from reconstructed frame $\recV$ of the previous frame. The focus of this paper lies on the highlighted inter coder.\label{Fig:Intro}}
	\vspace{-0.5cm}
\end{figure}

Since conditional inter frame coding is currently a widely researched area, we focus on one part of a hybrid coder, as schematically shown in Fig.~\ref{Fig:Intro}. The core of this inter coder is a coder, whose task it is to compress the current frame $\curV$ with the help of a temporal prediction $\predV$. In the following, we use the terms \emph{Residual Coder}, \emph{Conditional Coder}, and \emph{Conditional Residual Coder} to refer to specific embodiments of this coder, following specific coding paradigms, which are described throughout the paper. 

Residual inter coding is widely used due to practicality constraints and can be seen as a special case of the more general conditional inter coding. In traditional coders it is easier to model the distribution of the residual signal than to model the conditional distribution of an image given a predictor. In neural network based compression, where the distribution is modeled by neural networks and part of an end-to-end training process, it is possible to model the conditional distribution instead. Conditional inter coding was first proposed in~\cite{LadunePH2020_ModeNetModeSelection}. Since then, more and more state-of-the-art coders use conditional coding, for example DCVC~\cite{LiLL2021_DeepContextualVideo,ShengLL2022_TemporalContextMining,LiLL2022_HybridSpatialTemporal,LiLL2023_NeuralVideoCompression} or the purely transformer-based VCT~\cite{MentzerTM2022_VCTVideoCompression}.

One possibility to model conditional coding is the conditional autoencoder~\cite{SohnLY2015_LearningStructuredOutput}. A conditional autoencoder is a variant of an autoencoder, which has an additional input at both encoder and decoder, providing additional information about the signal. This way, the latent space can become smaller and only focuses on the new information. Early conditional autoencoders were used for example for handwriting recognition~\cite{HosoeYK2018_OfflineTextIndependent} or intra prediction~\cite{BrandSK2020_IntraFrameCoding}. 
\IEEEpubidadjcol

In a previous publication~\cite{BrandSK2022_BenefitsChallengesConditional}, we showed that under idealized conditions, conditional coding is always at least as good as residual coding and almost always better. However, we also argued that conditional coding is prone to containing information bottlenecks, which decrease the performance and therefore need to be avoided. 

Current conditional inter coders contain elements which can be shown to reduce bottlenecks. These methods are often computationally very  expensive and cumbersome. In this paper, we present the conditional residual coder, a hybrid between conditional and residual coding. We will show both in theory and on a practical example that this coder has a much higher robustness to bottlenecks and therefore allows a better performance at a lower complexity. 

The goal of this paper is to provide a thorough understanding of bottlenecks, how they affect the compression efficiency of conditional coders, and how they can be avoided using the conditional residual coding paradigm. This can guide the development of future development of deep-learning-based video coders. The main contributions of this paper are the following:
\begin{itemize}
	\item Deriving the conditional residual coder from an information theoretical analysis of information bottlenecks in conditional coders
	\item Proving that the conditional residual coder is at least as good as the conditional coder and less prone to bottlenecks
	\item An evaluation in trained inter frame coders to show the effect of bottlenecks in a conditional coder and the superior performance of the conditional residual coder
\end{itemize}

After Section II, where we give an overview over theoretical and practical work on conditional coding, the remainder of the paper is organized in two parts: In the first part consisting of Sections III - VI, we look at theoretical properties of conditional coders. Here, we derive the existence of bottlenecks and show their influence on the performance of conditional coders. Using information theoretical deliberations, we motivate the conditional residual coder and then prove that this concept is less prone to bottlenecks. We provide the analysis and derivations for the conditional residual coder both for lossless and lossy coding, thereby giving a complete description of the conditional residual coder.

In the second part consisting of Sections VII and VIII, we analyze implementations of conditional and conditional residual coders. Here, we focus on comparing high-level coding paradigms rather than low-level network architectures, since our results do not depend on specific network structures. The coders were designed such that the strength of the bottleneck can be steered. We present a simple method to measure and visualize bottlenecks and demonstrate that the conditional residual coder---different than the conditional coder---does not exhibit bottleneck effects anymore. The paper closes with a conclusion and an outlook in Section IX.


\section{Related Work}
\subsection{Theoretical Works on Conditional Coding} 
Conditional coding describes the process of compressing one source $X_1$ while knowing another source $X_0$. This is strongly related to joint compression of $X_1$ and $X_0$. In their paper~\cite{SlepianW1973_Noiselesscodingcorrelated}, Slepian and Wolf described a total of 16 different cases of conditioning both encoder and decoder and established what is now called the Slepian-Wolf bounds and Slepian-Wolf coding. The scenario which the authors describe is the compression and decompression of two dependent random variables $X_1$ and $X_0$, where the respective encoder and decoders may have additional knowledge about the respective other variable. In the case that both encoders and both decoders know the respective other variable, Slepian and Wolf showed that
\begin{equation}\label{Eq:SW1}
\rate_0 + \rate_1 \geq H(X_0,X_1)
\end{equation}
is feasible for the rates $\rate_0$ and $\rate_1$ of the signals $X_0$ and $X_1$, respectively. Here and in the following $H(X_0,X_1)$ denotes the joint (Shannon) entropy of $X_0$ and $X_1$. This result is intuitive, since one coder can compress both sources with $H(X_0,X_1)$.

If only the decoders have the additional knowledge, we need two further conditions
\begin{equation}\label{Eq:SW2}
\rate_0 \geq H(X_0|X_1)
\end{equation}
\begin{equation}\label{Eq:SW3}
\rate_1 \geq H(X_1|X_0).
\end{equation}
for feasibility, additionally to (\ref{Eq:SW1}). This case of independent encoders and dependent decoders is also known as distributed coding. 

In video compression, when we consider $X_0$ and $X_1$ as the sources of two consecutive frames, it is typically demanded that the frames are encoded and decoded sequentially. Therefore only the coder of one frame $X_1$ can have knowledge about the other frame. When conditional coding is used for such progressive inter coding, we have to consider a different scenario, which is also covered by Slepian and Wolf. One frame $X_0$ is coded independently, and another frame $X_1$ is coded conditionally. In this case the encoder of $X_1$ and the decoder of $X_1$ have additional knowledge about $X_0$. We get the feasible rate region for 
\begin{equation}\label{Eq:SW4}
\rate_0 \geq H(X_0)
\end{equation}
\begin{equation}\label{Eq:SW5}
\rate_1 \geq H(X_1|X_0)
\end{equation}
Curiously, in this coding scheme, it does not matter if the encoder makes use of $X_0$, as long as the decoder does. 


\subsection{Conditional Neural-Network-Based Video Compression}
A large number of publications on neural-network-based video compression~\cite{LuOX2018_DVCEndend,AgustssonMJ2020_ScaleSpaceFlow,HuLX2021_FVCNewFramework,HuLG2022_CoarseFineDeep,LinJZ2023_DMVCDecomposedMotion,YangTG2023_AdvancingLearnedVideo} use a classical residual coding approach. After motion compensation, the residual frame is computed by subtracting the prediction from the original frame to be transmitted. After the (lossy) transmission of the residual, the prediction frame is added to obtain the reconstructed frame. This procedure is by now a standard approach and was taken over from many generations of hybrid video compression standards~\cite{WiegandSB2003_OverviewH.264/AVCvideo,SullivanOH2012_OverviewHighEfficiency,BrossWY2021_OverviewVersatileVideo,MukherjeeBG2013_latestopensource,HanLM2021_TechnicalOverviewAV1} to end-to-end video compression.

In other works, conditional coding was used explicitly. This includes the works by Ladune \etal~\cite{LadunePH2020_ModeNetModeSelection, LadunePH2020_OpticalFlowMode,LaduneP2022_AIVCArtificialIntelligence}, where the authors propose a scheme with two encoders and one decoder. The main encoder compresses the image under the condition of knowing the motion compensated prediction signal, while a second prediction encoder transforms the prediction signal into a latent representation of the same spatial dimension as the main latent representation. Both of these representations are then concatenated and used to decode the image. Another key component of this work is a masking mechanism which enables a skip mode where part of the prediction frame is directly copied into the reconstructed frame, similar to the behavior of skip blocks in, e.g., HEVC~\cite{SullivanOH2012_OverviewHighEfficiency}. 

Recently more conditional coding schemes were proposed, like DCVC~\cite{LiLL2021_DeepContextualVideo} and its successors\cite{ShengLL2022_TemporalContextMining,LiLL2022_HybridSpatialTemporal,LiLL2023_NeuralVideoCompression}. Here, the authors also use a network to pre-process the prediction signal, however, both encoder and decoder receive the same signal at full pixel resolution. At the decoder, this requires an additional processing step at pixel resolution, where the reconstructed residual information and the prediction information are fused to obtain the reconstructed frame. Additionally, the context model is also conditioned on the prediction signal.

In~\cite{BrandSK2022_PFrameCoding}, we proposed another kind of conditional coding, which relies on generalized difference and sum operators. This way, the overall coding approach is still conditional, however, with a significantly lower complexity, since the additional encoder is not needed. The methods proposed by Hu~\etal in \cite{HuLX2021_FVCNewFramework} and Chen~\etal in~\cite{ChenWL2023_CompactTemporalTrajectory} lie somewhat in between conditional and residual compression. There, the residual is computed and transmitted in a feature space instead of the pixel space. On a high level, this qualifies as a conditional coder, while it still functions as a residual coder in the latent space. 

In~\cite{MentzerTM2022_VCTVideoCompression}, Mentzer \etal propose an approach which differs from most other schemes in that they do not use motion compensation but instead use a transformer along the temporal axis to predict priors for subsequent frames. Their approach can also be classified as a conditional coder since additional information is incorporated directly in both the encoder and decoder.

\vspace{-0.1cm}
\section{Conditional Coding and Residual Coding}
A conditional coder is an inter coder which codes one frame of a sequence under the condition of knowing a prediction frame. In contrast, a residual coder compresses the difference between a frame and its prediction. In the following, we always assume that both encoder and decoder know the prediction signal $\predV$. This signal was computed from a previously transmitted frame, possibly using motion information transmitted over a side channel. However, the following computations and deliberations are independent of the precise way the prediction signal is generated, as long as it is available at both encoder and decoder.

\begin{figure}
	\centering
	\begin{tabular}{cc}
		\includegraphics{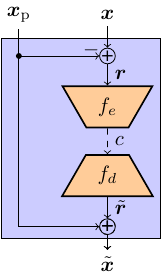}&
		\includegraphics{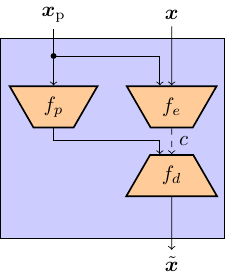}
		\\
		(a) Residual coder & (b) Conditional coder
	\end{tabular}
	\caption{Schematic diagram of a residual (a) and a conditional coder (b). Dashed lines represent signals which are transmitted. The blue frame indicates the position of this block in an inter coder, as shown in Fig.~\ref{Fig:Intro}.\label{Fig:CondVsResConcept}}
	\vspace{-0.5cm}
\end{figure}

To show the fundamental difference between residual coding and conditional coding, Fig.~\ref{Fig:CondVsResConcept} shows block diagrams of both coders. In the diagrams, $\curV$ denotes the frame to be transmitted, $\predV$ the prediction frame, and $\recV$ the reconstructed frame. The prediction frame is typically generated using motion compensation with a motion vector field which is transmitted as side information. $\intermedBefore$ denotes the signal which is sent over the channel. Additionally, in residual coding the residual frame is denoted as $\resV$ and the reconstructed residual as $\recresV$.


The residual coder uses the prediction signal (which is equally available at both encoder and decoder) to compute the residual signal via subtraction. From this residual signal, the bitstream $\intermedBefore$ is created. The reconstructed residual $\recresV$ is then derived from $\intermedBefore$. Finally, to obtain the reconstructed frame, the prediction frame $\predV$ needs to be added to $\recresV$.

In contrast to residual coding, conditional coding uses a different concept. The bitstream $\intermedBefore$ is directly generated from $\curV$ and $\predV$. After transmission, the reconstructed signal $\recV$ is generated from the intermediate signal $\intermedBefore$ on the decoder side and the prediction signal $\predV$. Many realizations of conditional coders, e.g.,~\cite{LadunePH2020_ModeNetModeSelection,LadunePH2020_OpticalFlowMode,LaduneP2022_AIVCArtificialIntelligence,LiLL2021_DeepContextualVideo,ShengLL2022_TemporalContextMining,LiLL2022_HybridSpatialTemporal} include a second network $\funcPredEnc$, which transforms the prediction signal into a feature domain before it is used for decoding. This function is used to match the spatial dimensions of the latent space or to extract important features from the prediction signal. The motivation for this block arises from the fact that the latent space contains the transmitted high-level features. This representation is incompatible with the pixel domain, so the prediction signal is transformed in a compatible domain before combining it with the latent representation. In~\cite{LiLL2021_DeepContextualVideo,ShengLL2022_TemporalContextMining,LiLL2022_HybridSpatialTemporal}, this block is called temporal context mining and is also used to generate additional input to the encoder. For simplicity, in the following, we only consider the case where the encoder receives the original prediction signal, since Slepian and Wolf~\cite{SlepianW1973_Noiselesscodingcorrelated} showed that, at least for the lossless case, the knowledge of the decoder determines the feasible rate, as shown in (\ref{Eq:SW5}). Also note that the network~$\funcPredEnc$ is not conceptually necessary for a conditional coder. In this case, the feature transform would still have to happen and would be implicitly performed in the decoder.

In practice, the functions the conditional coder uses to compute $\intermedBefore = f_e(\curV, \predV)$ and $\recV = f_d(\intermedBefore, f_p(\predV))$ are modeled with neural networks, and often with some kind of autoencoder as for example in~\cite{LadunePH2020_ModeNetModeSelection, LadunePH2020_OpticalFlowMode,LiLL2021_DeepContextualVideo,MentzerTM2022_VCTVideoCompression}, though other approaches exist~\cite{HoCC2022_CANFVCConditional}\cite{BrandSK2022_PFrameCoding}.

To summarize, in the residual coder, the reconstructed frame $\recV$ is computed as 
\begin{equation}
\recV = \predV + \funcDec\left(\funcEnc\left(\curV-\predV\right)\right) \label{Eq:RecResCoder}
\end{equation}
The required average rate to transmit the signal is given by the entropy of the residual signal 
\begin{equation}
\rate_\mathrm{Res} = H(\resD),
\end{equation}
for lossless transmission, where $\resD$ is the source emitting \mbox{$\resV=\curV-\predV$}. 
In the conditional coder, the reconstructed signal is computed with
\begin{equation}
\recV = \funcDec\left(\funcPredEnc\left(\predV\right),\funcEnc\left(\predV,\curV\right)\right), \label{Eq:RecCondCoder}
\end{equation}
which requires an average rate of
\begin{equation}
\rate_\mathrm{Cond,ideal} = H(\curD|\predD),
\end{equation}
where $\curD$ and $\predD$ are the sources of $\curV$ and $\predV$, respectively.

In lossy compression, the rate-distortion functions of the residual and conditional coder are given by
\begin{equation}
\RateRes(\dist) = \min_{p\in\PSet(\dist)}I(\resD;\recresD)\label{Eq:RDEqResidual}
\end{equation}
and
\begin{equation}
\RateCond(\dist) = \min_{p\in\PSet(\dist)}I(\curD;\recD|\predD)\label{Eq:RDEqConditional},
\end{equation}
respectively. Here, $\PSet(\dist)$ is the set of all probability functions $p(\curV,\predV,\recV)$, for which the average distortion does not exceed $\dist$ and where the marginal distribution $\sum_{\recV}p(\curV,\predV, \recV) = p_\mathrm{m}(\curV,\predV)$
is fixed. Note that $\resV$ and $\recresV$ can be computed from $\curV$, $\recV$ and $\predV$ and are therefore not explicitly mentioned in the probability distribution.

It is well-known that conditional coders can achieve superior performance in compression efficiency over a residual coder. In~\cite{BrandSK2022_BenefitsChallengesConditional}, we showed that 
\begin{equation}
\rate_\mathrm{Res} = H(\resD) = H(\curD|\predD) + I(\predD;\resD) \geq \rate_\mathrm{Cond,ideal}. \label{Eq:Main}
\end{equation}
Following similar steps as in \cite{BrandSK2022_BenefitsChallengesConditional}, we can show  
\begin{equation}
I(\resD;\recresD)=I(\curD;\recD|\predD) + \underbrace{I(\predD;\resD) - I(\predD;\resD|\recresD)}_{\geq 0}, \label{Eq:MainLossy}
\end{equation}
where the last term is greater or equal to zero since $I(\predD;\resD) \geq I(\predD;\resD|\recresD)$. Since this holds for all distributions, this implies
\begin{equation}
\RateRes(\dist)\geq\RateCond(\dist).\label{Eq:MainLossyX}
\end{equation}
A proof for this statement can be found in the appendix of this paper.
This shows that both in lossless and lossy coding, conditional coding can be superior to residual coding.

\section{Information Bottlenecks}
\subsection{Occurrence of Bottlenecks}
\label{sec:information-bottlenecks}
To understand how bottlenecks occur in a conditional coder, we look at the extreme case of perfect prediction, i.e., when $\predV=\curV$. In this case, no residual information has to be transmitted. In a residual coder, this would result in $\recresV=0$ which would then, according to (\ref{Eq:RecResCoder}), result in $\recV=\predV=\curV$. For a conditional coder, however, the prediction signal is part of the processing chain. When we do not need to transmit any residual information, according to (\ref{Eq:RecCondCoder}), we obtain
\begin{equation}
\recV = \funcDec\left(\funcPredEnc\left(\predV\right),\funcEnc\left(\predV,\curV\right)\right) = \bar\funcDec\left(\funcPredEnc\left(\predV\right)\right)= \bar\funcDec\left(\funcPredEnc\left(\curV\right)\right), \label{Eq:BNMotivation}
\end{equation}
where $\bar\funcDec\left(\cdot\right)$ denotes the function $\funcDec\left(\cdot, \funcEnc\left(\predV,\curV\right)\right)$, when the second argument is constant because no information is transmitted. 

Before reconstructing $\recV$, the prediction signal $\predV$ has to pass through two functions, which process the signal. This has two implications: First according to the data processing theorem~\cite{ViterbiO1979_PrinciplesDigitalCommunication}, processing of the data can only reduce the information content from $\predV$ which can be used to reconstruct $\recV$. Second, as we can see from (\ref{Eq:BNMotivation}), the system $\funcPredEnc$ and $\bar\funcDec$ form an autoencoder-like structure between $\predV$ and $\recV$. This structure needs the capacity to represent an entire image. In order to be able to use all information from $\predV$ to reconstruct $\recV$, this system needs to be lossless, which is not given in current conditional coding architectures. Due to these effects, it is not guaranteed that $\recV$ can be reconstructed accurately. This is in contrast to the residual coder. Here, the prediction signal is taken over directly without processing, so no bottleneck is present between $\predV$ and $\recV$. 

Another way to think about this is that the information content of $\recV$ comes from two paths: the prediction path, which connects $\predV$ and $\recV$ and the transmission path, which contains the transmission over the channel. The transmission path definitely contains a bottleneck in both cases, namely the channel. How much information is passed through this bottleneck determines the rate. For the residual coder, the prediction path is guaranteed to be bottleneck-free, since no information is lost and the sum is a reversible linear operator. For the conditional coder, the prediction path may contain a bottleneck. Due to insufficient network capacity, or simply due to the fact that neural networks are generally not reversible it is not guaranteed that all information from $\predV$ can be passed on to $\recV$. This gap has to be compensated by the transmission path, which thereby increases the necessary rate. 

In order to describe the bottleneck, we introduce \mbox{$\predBNV=f(\predV)$}, which is a processed version of $\predV$, and therefore by data processing theorem~\cite{ViterbiO1979_PrinciplesDigitalCommunication} satisfies
\begin{equation}
I(\predBND;\curD) \leq I(\predD;\curD),
\end{equation}
where $\predBND$ denotes the source of $\predBNV$ and $I(\cdot;\cdot)$ denotes the mutual information of two sources.

We now use the following model for bottlenecks. Since the reconstructed signal does not directly depend on $\predV$ but rather on a processed version $\predBNV$ of $\predV$, the results from Slepian and Wolf~\cite{SlepianW1973_Noiselesscodingcorrelated} assert that the required rate for transmission is given by
\begin{equation}
\rate_\mathrm{Cond} = H(\curD|\predBND) \geq H(\curD|\predD).
\end{equation}
The inequality follows again from the data processing theorem. Furthermore, the rate-distortion function for a real conditional coder considering bottlenecks is given by
\begin{equation}
\RateCondBN(\dist) = \min_{p\in\PSetBN(\dist)}I(\curD;\recD|\predBND). \label{Eq:RDEqConditionalBN}
\end{equation}
Here, $\PSetBN$ is equivalent to $\PSet$, except the probability distributions also describe $\predBNV$. Note that we assume that the marginal distributions of $\curV$ and $\predV$ are the same in both sets. Since $I(\curD;\recD|\predBND)\geq I(\curD;\recD|\predD)$ holds for all distributions, this implies
\begin{equation}
\RateCondBN(\dist) \geq \RateCond(\dist)
\end{equation} 

Note that the severity of bottlenecks highly depends on the implementation details of the coder. Conditional coders in literature feature concepts such as additionally transmitted skip modes~\cite{LadunePH2020_ModeNetModeSelection}, additional enhancement layers~\cite{LiLD2020_SpatialChannelContext}, or adaptive switching between conditional and residual coders~\cite{BrandSK2022_PFrameCoding}. We will show later that effects of the bottleneck can also be reduced by strongly increasing the capacity of the networks in the conditional path. This however also increases the complexity with all associated disadvantages, like increased runtime and memory footprint and an increased danger of overfitting.

\vspace{-0.3cm}
\subsection{Influence of Bottlenecks on the Conditional Coder}
In the following, we compute the effects of bottlenecks in the prediction path, first for lossless then for lossy coding. Since we model the bottleneck with a general function \mbox{$\predBNV = f\left(\predV\right)$}, where $\predBNV$ is a degraded version of $\predV$, it is clear from the data processing theorem that
\begin{equation}
I(\curD;\predD) \ge I(\curD;\predBND)
\label{Eq:Entropy}
\end{equation}
and thereby
\begin{equation}
H(\curD|\predD) \le H(\curD|\predBND)
\label{Eq:CondEntropy}
\end{equation}
holds. From Slepian and Wolf~\cite{SlepianW1973_Noiselesscodingcorrelated}, it follows that, when considering the presence of bottlenecks, $H(\curD|\predBND)$ acts as lower bound for the rate. The question now is whether this conditional entropy is also smaller or equal to $H(\resD)$.

To answer this question, it can be shown that 
\begin{equation}
H(\curD|\predD) = H(\curD|\predBND) - I(\curD;\predD|\predBND).
\label{Eq:CondEntropy2}
\end{equation}
The result fits the intuition, since $I(\curD;\predD|\predBND)$ is the amount of information $\predV$ additionally provides about $\curV$ and is not covered by $\predBNV$. Therefore this information has to be transmitted additionally. Eq. (\ref{Eq:CondEntropy2}) can be proven by expanding the conditional mutual information:

\begin{subequations}
	\begin{align}
	I(\curD;&\predD|\predBND) = I(\curD;\predD,\predBND) - I(\curD;\predBND) \\
	&= I(\curD;\predD) - I(\curD;\predBND)~\label{Eq:IA} \\
	&= H(\curD) - H(\curD|\predD) - H(\curD) + H(\curD|\predBND)\\
	&= H(\curD|\predBND) - H(\curD|\predD).
	\end{align}
\end{subequations}
In (\ref{Eq:IA}), we exploited that $\predBNV$ does not contain any information about $\curV$ which is not also present in $\predV$, which implies $I(\curD;\predD,\predBND) = I(\curD;\predD)$. 

Plugging this result into (\ref{Eq:Main}), we obtain
\begin{equation}
H(\resD) = H(\curD|\predBND) - I(\curD;\predD|\predBND) + I(\predD;\resD).
\label{Eq:Main2}
\end{equation}
Since $I(\curD;\predD|\predBND)$ is non-negative, $H(\resD)\geq H(\curD|\predBND)$ does not necessarily hold. Depending on how much information is lost during the bottleneck $f$, the rate might therefore increase for a conditional coder. 

Following a similar argument, we can show that 
\begin{equation}
I(\curD;\recD|\predBND) = I(\curD;\recD|\predD)+ I(\curD;\predD|\predBND) - I(\curD;\predD|\recD,\predBND)
\end{equation}
Plugging this into (\ref{Eq:MainLossy}), we obtain
\begin{equation}
\begin{split}
I(\resD;\recresD)&=I(\curD;\recD|\predBND) + \underbrace{I(\predD;\resD) - I(\predD;\resD|\recresD)}_{\geq 0}\\ &- \underbrace{\left(I(\curD;\predD|\predBND) - I(\curD;\predD|\recD,\predBND)\right)}_{\geq 0},
\end{split}
\end{equation}
Here, we see that we run into a similar problem that it is not possible to guarantee any more that $I(\curD;\recD|\predBND)\leq I(\resD;\recresD)$ and thereby $\RateCondBN(\dist)\leq\RateRes(\dist)$. When taking bottlenecks into account, conditional coding might hence no longer be better than residual coding due to losses imposed by the network structure.

\vspace{-0.2cm}
\section{Conditional Residual Coding}
\label{sec:CondResCoding}

\begin{figure}
	\centering
	\includegraphics{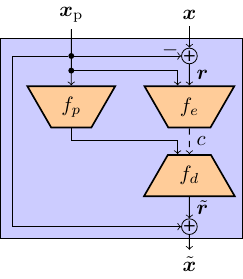}
	\vspace{-0.3cm}
	\caption{Schematic of the conditional residual coder.\label{Fig:CondResSchematic}}
	\vspace{-0.5cm}
\end{figure}

In the previous section, we took a closer look at conditional coding and showed that $H(\resD) \geq H(\curD|\predD)$ holds in general. However, we also showed that when we take the information bottlenecks between the prediction and the reconstructed signal into account, the overall performance may be below the residual coder. In this section, we propose the conditional residual coding paradigm for inter frames, which compresses the residual $\resV$ under the condition of knowing $\predV$. The basic structure is shown in Fig.~\ref{Fig:CondResSchematic}. Comparing the coder with the conditional coder in Fig.~\ref{Fig:CondVsResConcept}(b), we see that only a residual structure was added to the coder. This also means that this extension is practically for free since only one addition and one subtraction are required per pixel. We furthermore see that the conditional residual coder includes a bottleneck-free path from $\predV$ to $\recV$, which already suggests larger robustness. In lossless coding, the required rate is given by $H(\resD|\predD)$ or $H(\resD|\predBND)$, without and with considering bottlenecks, respectively. The rate distortion functions are given by:
\begin{align}
\RateResCond(\dist) =& \min_{p\in\PSetBN(\dist)}I(\resD;\recresD|\predD)\label{Eq:RDEqResConditional}\\
\RateResCondBN(\dist) =& \min_{p\in\PSetBN(\dist)}I(\resD;\recresD|\predBND)\label{Eq:RDEqResConditionalBN}
\end{align}

In this section, we first motivate the concept from basic information theoretical properties and prove three properties:
\begin{enumerate}
	\item The conditional residual coder is always better or equal to the residual coder, even with bottlenecks.
	\item When neglecting bottleneck effects, the conditional residual coder is equivalent to the conditional coder.
	\item When taking bottlenecks into account, the conditional residual coder is never worse than the conditional coder as long as $H(\recresD)<H(\recD)$ holds.
\end{enumerate}

\subsection{Ideal Lossless Conditional Residual Coder}
The motivation for a conditional residual coder becomes clear when looking at the following inequality which follows from basic identities of information theory and the data processing theorem:
\begin{equation}
H(\resD) \geq H(\resD|\predBND) \geq H(\resD|\predD).\label{Eq:BottlenckIneqRes}
\end{equation}
This asserts that $H(\resD|\predBND)$ is never larger than $H(\resD)$ and thereby never worse than the baseline of residual coding. This proves the first property of the conditional residual coder. Note that this does not mean that there are no bottleneck effects, however, the effect is bounded in a way that the corresponding coder can never be worse than the residual coder. The intuitive motivation for such a coder design is that the residual contains much less information than the frame itself. Therefore we need to pass on less information through the bottleneck, allowing for much tighter bottlenecks.

To prove the second statement, we want to show that without considering bottlenecks, transmitting $\curV$ is equivalent to transmitting $\resV$, or that 
\begin{equation}
H(\resD|\predD) = H(\curD|\predD).\label{Eq:Symmetry}
\end{equation}
This can easily be shown when considering 
\begin{equation}
\resV = \curV - \predV. \label{Eq:FunRes}
\end{equation}
Since $\curV$ is fully determined by $\resV$ and $\predV$ and $\resV$ is fully determined by $\curV$ and $\predV$, we can follow
\begin{equation}
H(\curD|\resD,\predD) = 0 \label{Eq:xentropy}
\end{equation}
\begin{equation}
H(\resD|\curD,\predD) = 0. \label{Eq:rentropy}
\end{equation}
Equation (\ref{Eq:Symmetry}) follows directly from there with
\begin{subequations}
	\begin{align}
	H(\resD|\predD) &= H(\resD|\predD) + \underbrace{H(\curD|\resD,\predD)}_{~~~~~~=0~(\ref{Eq:xentropy})}\\
	&=H(\curD,\resD|\predD)\\
	&=H(\curD|\predD) + \underbrace{H(\resD|\curD,\predD)}_{~~~~~~=0~(\ref{Eq:rentropy})}\\
	&= H(\curD|\predD),
	\end{align}
\end{subequations}
which shows that, without considering bottlenecks, conditional residual coding is equivalent to the conditional coder, thus proving the second property. 

Up to this point, we showed that the conditional residual coder is equal to the conditional coder without considering bottlenecks and never worse than the residual coder, even with bottlenecks. This however does not yet show that the conditional residual coder generally performs better than the conditional coder for any bottleneck. We examine this property in the next subsection.
\subsection{Considering Bottlenecks in the Lossless Coder}

In order to show that conditional residual coding in fact is less sensitive to bottlenecks, we include one further assumption in our framework:  
\begin{equation}\label{Eq:Assumption1}
H(\resD) < H(\curD)
\end{equation}
This assumes that the prediction is (on average) a good prediction, and produces a residual signal which can be transmitted better than the original signal. In any realistic coding scenario, this is the minimum requirement for any prediction. E.g., in VVC, temporally predicted inter frames require about five times less rate than intra frames~\cite{BrossWY2021_OverviewVersatileVideo}. This shows that the temporal residual requires less rate than an image. This assumption is therefore well justified.\hle

We start answering this question by looking at the extreme cases: In the case of no bottleneck (i.e., an infinitely wide bottleneck), $\predBNV = \predV$ holds, and we already showed that both approaches perform equally here. On the other hand, when the bottleneck is so tight that $\predBNV$ does not carry any information, we have $H(\curD|\predBND) = H(\curD)$ and $H(\resD|\predBND) = H(\resD)$. Since we assume that $H(\resD) < H(\curD)$ holds in a realistic case, transmitting $\resV$ is more efficient in this case. Considering the corner cases, the conjecture that $H(\resD|\predBND) \leq H(\curD|\predBND)$ holds in general does not seem too far fetched, nevertheless, we will show it in the following.

To show the effect of bottlenecks, we need to assure that the bottlenecks in the conditional coder and in the conditional residual coder are equally tight. To that end, we need to quantify the strength of a bottleneck. Here, we model bottlenecks such that they are able to pass on only $\BN$ bits of information. We can then write\footnote{Why do we model the bottleneck on $I(\curD;\predBND)$ and not on $H(\predBND)$? The quantity $H(\predBND)$ is actually irrelevant, since the only information which needs to be passed to $\recV$ is the information about $\curV$ contained in $\predBNV$, which is measured by $I(\curD;\predBND)$.}
\begin{equation}
I(\curD;\predBND) = \begin{cases}
\BN & ~~~\mathrm{if}~I(\curD;\predD)\geq\BN \\
I(\curD;\predD)& ~~~\mathrm{if}~I(\curD;\predD)<\BN
\end{cases}\label{Eq:BNModel}
\end{equation}
and 
\begin{equation}
I(\resD;\predBND) = \begin{cases}
\BN & ~~~\mathrm{if}~I(\resD;\predD)\geq\BN \\
I(\resD;\predD)& ~~~\mathrm{if}~I(\resD;\predD)<\BN
\end{cases}.\label{Eq:BNModel2}
\end{equation}
In the experiments in Section~\ref{Sec:Exp}, we will justify this simplified model of a bottleneck.
We can furthermore show that 
\begin{subequations}
	\label{Eq:BFull}
	\begin{align}
	H(\curD) - H(\resD) &= H(\curD) - H(\curD|\predD) - H(\resD) + H(\resD|\predD)\label{Eq:B}\\
	&= I(\curD;\predD) - I(\resD;\predD),
	\end{align}
\end{subequations}
where (\ref{Eq:B}) uses the symmetry from (\ref{Eq:Symmetry}), which asserts that $H(\resD|\predD) - H(\curD|\predD) = 0$. This shows together with the assumption (\ref{Eq:Assumption1}) that
\begin{equation}
I(\resD;\predD) \leq I(\curD;\predD) \label{Eq:Follow}
\end{equation}
holds. When considering the bottlenecks, we therefore have to look at three cases:
\begin{equation}
I(\resD;\predD) \leq I(\curD;\predD) \leq \BN \label{Eq:Case1}
\end{equation}
\begin{equation}
I(\resD;\predD) \leq \BN \leq I(\curD;\predD)  \label{Eq:Case2}
\end{equation}
\begin{equation}
\BN \leq I(\resD;\predD) \leq I(\curD;\predD)  \label{Eq:Case3}
\end{equation}

To show that conditional residual coding is always better or equal to conditional coding, we need to show that 
\begin{subequations}
	\begin{align}
		\begin{split}
		H&(\curD|\predBND)\!- \!H(\resD|\predBND) \\ &=[H(\curD)\!-\!I(\curD;\predBND)] - [H(\resD)\!-\!I(\resD;\predBND)]
		\end{split}
	 \\
	&= I(\curD;\predD) - I(\resD;\predD) - [I(\curD;\predBND) - I(\resD;\predBND)]\label{Eq:C}
	\end{align}
	\label{Eq:CFull}
\end{subequations}
is larger or equal to zero. Here, (\ref{Eq:C}) follows from ($\ref{Eq:BFull}$)

In the first case (\ref{Eq:Case1}), we see from (\ref{Eq:BNModel}) that $I(\curD;\predBND) = I(\curD;\predD)$ and $I(\resD;\predBND) = I(\resD;\predD)$ hold. Plugging this into (\ref{Eq:CFull}), we obtain
\begin{equation}
\begin{split}
&H(\curD|\predBND) - H(\resD|\predBND) \\&= I(\curD;\predD) - I(\resD;\predD) - (I(\curD;\predD) - I(\resD;\predD)) = 0.
\end{split}
\end{equation}
This is not surprising, since the bottleneck is not in effect in this case, thereby conditional and conditional residual coding are equivalent, as stated in (\ref{Eq:Symmetry}).

In the second case (\ref{Eq:Case2}) with a tighter bottleneck, the bottleneck first causes $I(\curD;\predBND) = \BN$. Thereby we obtain
\begin{subequations}
	\begin{align}
	\begin{split}
	H(\curD&|\predBND) - H(\resD|\predBND) \\&= I(\curD;\predD) - I(\resD;\predD) - (\BN - I(\resD;\predD))
	\end{split}
	 \\&= I(\curD;\predD) - \BN \geq 0.
	\end{align}
	
\end{subequations}
That the last expression is greater or equal to zero follows directly from the definition of the second case (\ref{Eq:Case2}).

In the third case (\ref{Eq:Case3}), $I(\curD;\predBND) = I(\resD;\predBND) = \BN$ holds and thereby
\begin{subequations}
	\begin{align}
	\begin{split}
	H(\curD&|\predBND) - H(\resD|\predBND) \\&= I(\curD;\predD) - I(\resD;\predD) - (\BN - \BN) 
	\end{split}
	\\&= I(\curD;\predD) - I(\resD;\predD) > 0,
	\end{align}
\end{subequations}
where the final inequality follows from (\ref{Eq:Follow}).

We see that for arbitrarily tight bottlenecks, a conditional residual coder always performs at least as good as a conditional coder, as long as the basic condition \mbox{$H(\resD)<H(\curD)$} is fulfilled. That way, we proved the third statement and have shown that the lossless conditional residual coder always performs better or equal to both the conditional and the residual coder.

\subsection{Lossy Conditional Residual Coder}
In the following, we prove the analogous statements for the lossy case. We again show the proofs for the three statements from above individually. The proofs follow similar steps as the corresponding proofs for the lossless case.

The first property can be easily shown since
\begin{equation}
I(\resD;\recresD|\predBND) \leq I(\resD;\recresD)
\end{equation}
holds in general and therefore $\RateResCondBN(\dist) \leq \RateRes(\dist)$.

We can show that without considering a bottleneck, the lossy conditional and the lossy conditional residual coder are equivalent by writing the mutual information as 
\begin{subequations}
	\label{Eq:SymmetryLossy}
	\begin{align}
	I(\resD&;\recresD|\predD) = H(\resD|\predD) + H(\recresD|\predD) - H(\resD, \recresD|\predD)\\
	&= H(\curD|\predD) + H(\recD|\predD) - H(\curD, \recD|\predD) \\&= I(\curD;\recD|\predD).
	\end{align}
\end{subequations}
The identities for the conditional entropies can be derived analogously to the identity $H(\resD|\predD) = H(\curD|\predD)$ in (\ref{Eq:Symmetry}). This symmetry can thereby be extended to the lossy case in a straightforward way.

To prove the third property, we use the same strategy as in the lossless case by looking at three different cases for the bottleneck. As a prerequisite, we need to prove several lemmas. 

Note that the bottleneck now affects the mutual informations $I(\recresD;\predD)$ and $I(\recD;\predD)$. Analogously to (\ref{Eq:BFull}), we can derive
\begin{equation}
I(\recD;\predD) - I(\recresD;\predD) = H(\recD) - H(\recresD),
\end{equation}
which shows that $I(\recD;\predD) \geq I(\recresD;\predD)$ if $H(\recD) \geq H(\recresD)$.

We can furthermore derive that 
\begin{subequations}
	\begin{align}
	I&(\curD;\recD) - I(\resD;\recresD) = I(\curD;\recD) - I(\resD;\recresD)  \nonumber\\&+ \underbrace{I(\resD;\recresD|\predD) - I(\curD;\recD|\predD)}_{~~~~~~=0 ~(\ref{Eq:SymmetryLossy})} \\
	&=I(\recD;\predD) - I(\recD;\predD|\curD) - I(\recresD;\predD) + I(\recresD;\predD|\resD)
	\end{align}
\end{subequations}

Note that in the equivalent result from (\ref{Eq:BFull}), the two conditional mutual information terms did not appear. We can write 
\begin{equation}
I(\recD;\predD|\curD) = I(\recD;\predBND|\curD) + I(\recD;\predD|\curD,\predBND).
\end{equation}
This shows that the mutual information $I(\recD;\predD|\curD)$ can be split into two parts: First $I(\recD;\predBND|\curD)$ which models the information from $\predV$ which is passed to $\recV$ through the prediction path. Second, $ I(\recD;\predD|\curD,\predBND)$ which can not be passed through the prediction path, since it measures the mutual information of $\recV$ and $\predV$ if $\predBNV$ is known, therefore adding to the information available through $\predBNV$. This information therefore has to be transmitted. However, an ideal coder would never transmit this information, since it also does not contain information about $\curV$, as it is conditioned on $\curD$. Any information which is described by $I(\recD;\predD|\curD,\predBND)$ can thereby not contribute to an accurate reconstruction of $\curV$ at the decoder, but still costs rate. We therefore conclude that in an optimal coder $I(\recD;\predD|\curD,\predBND)=0$ and thereby
\begin{equation}
I(\recD;\predD|\curD) = I(\recD;\predBND|\curD)
\end{equation}
holds. We can make the equivalent argument to show 
\begin{equation}
I(\recresD;\predD|\resD) = I(\recresD;\predBND|\resD).
\end{equation}

From this we can follow that 
\begin{align}
I&(\curD;\recD) - I(\resD;\recresD) \label{Eq:CondResLossyA}\\&= I(\recD;\predD) - I(\recresD;\predD) - I(\recD;\predBND|\curD) + I(\recresD;\predBND|\resD) \notag
\end{align}

We can then continue to express the gain of conditional residual coding in terms of bottleneck-relevant mutual informations:
\begin{subequations}
	\begin{align}
	I(\curD;& \recD|\predBND) - I(\resD; \recresD|\predBND)\nonumber\\=& I(\curD;\recD) - I(\resD;\recresD) - I(\recD;\predBND) \nonumber\\&+I(\recresD;\predBND) + I(\recD;\predBND|\curD) - I(\recresD;\predBND|\resD)\\
	=& I(\recD;\predD) - I(\recresD;\predD) - I(\recD;\predBND) + I(\recresD;\predBND)
	\end{align}
\end{subequations}
This expression is equivalent to (\ref{Eq:CFull}). From this point, we can now follow the same proof as above by considering the following three cases
\begin{equation}
I(\recresD;\predD) \leq I(\recD;\predD) \leq \BN
\end{equation}\vspace{-0.6cm}
\begin{equation}
I(\recresD;\predD) \leq \BN \leq I(\recD;\predD) 
\end{equation}\vspace{-0.6cm}
\begin{equation}
\BN \leq I(\recresD;\predD) \leq I(\recD;\predD). 
\end{equation} 
We can then show, equivalently as for the lossless case, that $I(\recD;\predD) - I(\recresD;\predD) - I(\recD;\predBND) + I(\recresD;\predBND) \geq 0$ for each case. This implies \begin{equation}
I(\curD; \recD|\predBND) \geq I(\resD; \recresD|\predBND).
\end{equation}
Since this holds for all sensible distributions, this finishes the proof that 
\begin{subequations}
	\begin{align}
	\RateResCondBN(\dist) &= \min_{p\in\PSetBN(\dist)}I(\resD;\recresD|\predBND)\\
	&\leq \min_{p\in\PSetBN(\dist)}I(\curD;\recD|\predBND) = \RateCondBN(\dist)
	\end{align}
\end{subequations}
and thereby shows that conditional residual coding is always as least as good as conditional coding and never worse than residual coding. Conditional residual coding therefore combines both the best of residual and conditional coding.

\section{Model-Based Validation for Lossless Coding}
In the following, we analyze the performance of conditional and conditional residual coding on a simple image model. The main purpose of this experiment is to gain insight into the bottleneck effects. We will use these insights in the subsequent section to analyze a trained inter coder.

\subsection{Image Model}
Since there is no accurate probability model for an entire image, it is not possible to compute the entropy of an image or the conditional entropy of one image given another. Furthermore, there are no established models connecting the prediction frame $\predV$ and the current frame $\curV$. Since we are more interested in temporal than in spatial redundancy here, we limit our model to the transmission of one pixel $\cur$ under temporal prediction with the predictor $\pred$. Both $\cur$ and $\pred$ assume values in the 8 bit range from 0 to 255.

To model temporal redundancy, we consider an ideal motion compensation based predictor with the following properties: If there is a match in the reference frame, the predictor will find this match. If there is no match, as is the case, e.g., for occlusions, then we assume that the prediction will assume a random and uncorrelated value, i.e., the value for $\pred$ is then drawn from an iid uniform distribution. We denote the probability for an occlusion with $\probMVE$. We can therefore write the probability for $\pred$ given $\cur$ as:
\begin{equation}
\Prob(\pred|\cur) = \frac{1}{256}\cdot\probMVE + (1-\probMVE)\cdot\delta\left[\pred-\cur\right],
\end{equation}
with $\delta[\cdot]$ denoting the delta unit impulse.
Note that in this model we do not consider noisy images. In~\cite{BrandSK2022_BenefitsChallengesConditional}, we showed in an evaluation with a similar model that noise does not affect the gain of conditional coding but that a higher noise level reduces the influence of bottlenecks. The noise-free case is therefore the most challenging case for conditional coders, which is why we focus on this case in the following. 

Additionally, we also need a model for information bottlenecks in the prediction path. We model the bottlenecks using quantization, such that
\begin{equation}
\predBN = \left\lfloor\pred/\QuantStep\right\rceil\cdot\QuantStep,
\end{equation} 
with the rounding operation $\lfloor\cdot\rceil$ and the quantization step size $\QuantStep$, which constitutes the second model parameter. 

With this model, it is now possible to directly compute $H(\resD)$, $H(\curD|\predD)$, and $H(\curD|\predBND)$ under different parameter settings. We can thereby show the behavior of the different methods under controlled conditions.

\vspace{-0.2cm}
\subsection{Model Results}
\begin{figure*}
	\centering
	\begin{tabular}{cc}
		
		\multicolumn{2}{c}{
			\scalebox{0.85}{
				\includegraphics{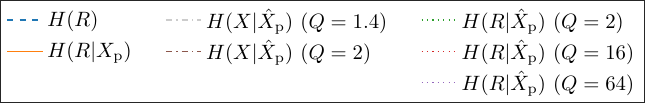}
			}
		}\\
		
		\includegraphics{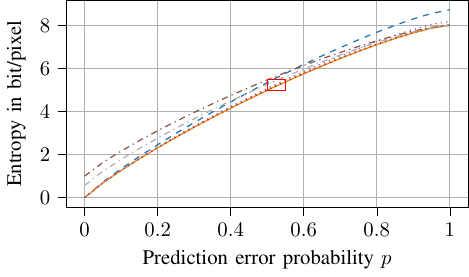}
		&
		\includegraphics{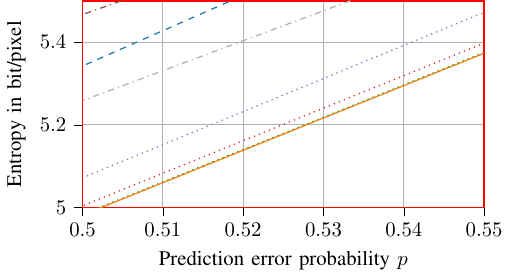}
		\\(a) Computed entropies&(b) Zoomed excerpt
	\end{tabular}
	\caption{Computed entropies of residual, conditional, and conditional residual coders when varying $\probMVE$. We show plots for different bottlenecks. As a comparison, we also show the result for a normal conditional coder. In (b), a zoomed excerpt is shown. The corresponding area is marked with a red frame in (a). Note that according to (\ref{Eq:Symmetry}), $H(\resD|\predD) = H(\curD|\predD)$.\label{Fig:EntropyRes}}
	\vspace{-0.5cm}
\end{figure*}

Fig.~\ref{Fig:EntropyRes} shows the computed entropies when the prediction error probability $\probMVE$ is varied. We note several effects: $H(\curD|\predD)$ is always below $H(\resD)$, so the ideal conditional coder is always better than the residual coder. This is in line with our theory. Note that the gap between them increases with $\probMVE$, i.e., with decreasing prediction quality. When $\cur$ and $\pred$ are uncorrelated, $\resD$ has an entropy of 8.72 bit/pixel yielding a gain of 0.72 bit/pixel

When we take bottlenecks into account, (dash-dotted lines), we see that especially for small $\probMVE$ the entropy $H(\curD|\predBND)$ is larger than $H(\resD)$. This shows that in this case the conditional coder performs worse than the residual coder. Bottleneck effects are more severe for small $\probMVE$, i.e., for high prediction quality. Only with decreasing quality, the conditional coder eventually becomes better than the residual coder again. Note that with $Q=2$ and $Q=1.4$, we chose relatively wide bottlenecks, only loosing 1 and 0.5 bit/pixel of information.

For the conditional residual coder (dotted lines), this looks completely different. For a bottleneck with $Q=2$, the curve is almost not distinguishable from $H(\curD|\predD)$. Even in the zoomed excerpt in Fig.~\ref{Fig:EntropyRes}(b), the curves almost coincide. Note the strong contrast to the $Q=2$ case in the conditional coder, which performs worse than the residual coder over a large range of $\probMVE$. We can tighten the bottleneck to $Q=64$, which corresponds to a loss of 6 bit/pixel in the prediction path and the performance is still clearly better than the residual coder. Here, we can see the advantage that $H(\resD|\predBND)$ is upper bounded by $H(\resD)$, as opposed to $H(\curD|\predBND)$, which is upper bounded by $H(\curD)$. 

\section{Analysis on Trained Lossy Coders}
In the previous section, we have analyzed the different coding paradigms on a simple image model. This allowed large control over the statistical properties of the image and the prediction signal and gave valuable insights about conditional coders. However, naturally a simple model can not cover all the complexity of natural images. In particular, our model did not consider spatial correlation, which can be exploited during compression. To analyze conditional and conditional residual coders also on real images, we show the performance of conditional coders within an inter coding setup in this section. To that end, we train a residual coder and several conditional coders, designed to have differently tight bottlenecks. In the following experiments, we will use lossy coders, which is the typical use-case for video coding. The experiments will confirm that conditional coders exhibit bottleneck effects and that, as predicted in the theoretical analysis, conditional residual coders are essentially free of such effects.

\subsection{Setup}\label{Sec:Exp}
\begin{figure*}
	\centering

	\begin{tabular}{ccc}
		\includegraphics{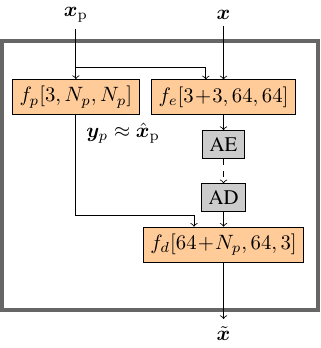}
		&
		\includegraphics{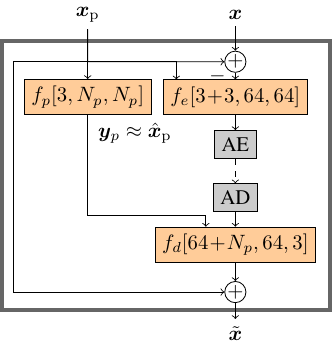}
		&
		\includegraphics{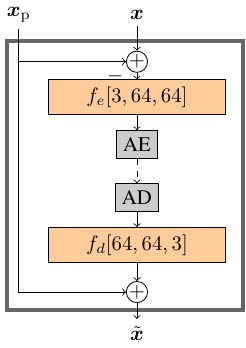}
		\\
		(a) Conditional coder~\cite{LadunePH2020_ModeNetModeSelection} & (b) Conditional residual coder & (c) Equivalent residual coder
	\end{tabular}
	\caption{Structure of a conditional (a), a conditional residual (b) and a residual coder (c). $\encoderNet$ and $\decoderNet$ denote the encoder and decoder part of an autoencoder, respectively. $\predChannels$ denotes the number of channels in the prediction encoder. All networks are convolutional networks, exactly as proposed in \cite{LadunePH2020_ModeNetModeSelection}. The three arguments denote the number of input channels, the number of intermediate channels  and the number of output channels. For the channels, the networks have a context model with hyperprior as proposed in~\cite{LadunePH2020_ModeNetModeSelection}.\label{Fig:CondCoderNet}}
	\vspace{-0.5cm}
\end{figure*}
We designed the following experiments to demonstrate the influence of bottlenecks on conditional and conditional residual coders. To that end, we examine \textit{CodecNet} as proposed in~\cite{LadunePH2020_ModeNetModeSelection} and an equivalent conditional residual coder. We choose CodecNet due to its simplicity and because the tightness of the bottleneck depends on the number of channels in the auxiliary latent space. We can thereby steer the tightness of the bottleneck by introducing the number of channels $\predChannels$ as a parameter and approximate the virtual variable $\predBNV$ with the latent representation $\latentPredV$ of $\predV$. That way, we can observe the influence of conditional and conditional residual coding with differently tight bottlenecks. The broad structure of CodecNet is shown in Fig.~\ref{Fig:CondCoderNet}(a). Note that CodecNet was originally proposed together with \textit{ModeNet}, which enables a skip mode bypassing the conditional coder to directly use the prediction signal in some areas. In the following, we do not use ModeNet since our focus is on comparing different residual coding structures. The influence of ModeNet on bottlenecks is further discussed in Section~\ref{Sec:RelToSoTA}.

To design the conditional residual coder for this experiment, we use the exact same components as for the conditional coder. The only difference is that we encode the difference $\resV$ instead of $\curV$. The structure of this coder can be seen in Fig.~\ref{Fig:CondCoderNet}(b) As a baseline, we use an equivalent residual coder, which was designed in the same way as the residual baseline in~\cite{LadunePH2020_ModeNetModeSelection}. Fig.~\ref{Fig:CondCoderNet}(c) shows the schematic of this coder. In particular, the number of channels in encoder and decoder is the same as for the conditional coder.

We used an MSE distortion function $\distMSE(\cdot;\cdot)$ and a rate estimation function $\rate(\cdot)$ to train the different coders on a joint rate-distortion loss:
\begin{equation}
\loss = \lagrange\distMSE(\curV;\recV) + \rate(\latent).
\end{equation}
We trained the network for 30 epochs on the CLIC 2020 P-frame coding training set~\cite{Mentzer_clic2020devkit}. We used the Adam~\cite{KingmaB2015_AdamMethodStochastic} algorithm with a learning rate of $10^{-4}$. to train four models for four rate points with \mbox{$\lagrange\in\left\{256,512,1024,2048\right\}$}. For motion compensation and motion transmission, we used pre-trained DVC~\cite{LuOX2018_DVCEndend} components, which are not updated during the training. There are two reasons for this strategy: Since the surrounding components are exactly the same for all tested models, all models get exactly the same prediction signals. We thereby obtain a fair comparison of the different inter coding methods. Furthermore, to demonstrate the bottleneck effects, we split the test set according to the achieved prediction quality. Having the same prediction signal for all methods, assures that the splits are the same for all cases and thereby the validity of the experiment.

We test the networks on the CLIC 2020 P-frame validation set~\cite{Mentzer_clic2020devkit}. The set consists of 13 classes with vastly differing content. Note that the set also contains extreme cases, including picture repetition, very noisy sequences and scene cuts. With this wide variety of content, we can test conditional coding under various conditions. For each class, we randomly\footnote{Of course the choices are the same for all test cases.} select 100 frame pairs from which we crop $512\!\times\!512$ patches, which are to be encoded. We compress the reference frame with the neural network based coder from~\cite{MinnenBT2018_JointAutoregressiveHierarchical}, as found in \textit{compressai}~\cite{BegaintRF2020_CompressAIPyTorchlibrary}. We generate a bitstream for each P-frame and measure the required bits per pixel and the reconstruction quality in terms of PSNR.

\subsection{Coding Results}
\begin{figure*}
	\centering
	\begin{tabular}{ccc}
		\includegraphics{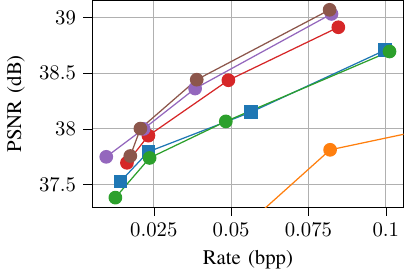}
		&
		\includegraphics{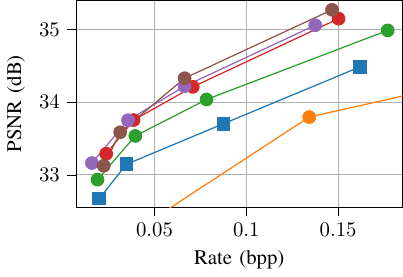}
		&
		\scalebox{0.85}{
			\includegraphics{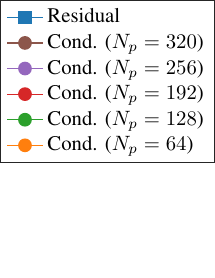}
		}
		\\
		~~~~~~~~~~~(a) Entire test set & ~~~~~~~~~~~~~~(b) $\predPSNR \leq 35\dB$
	\end{tabular}\vspace{-0.1cm}
	\caption{Rate-distortion curves for residual and conditional coder. In (a) the entire test set is used. In (b) only cases where the prediction PSNR does not exceed 35\dB.\label{Fig:ResultsCondVsDiff} \vspace{-0.1cm}}
	\vspace{-0.5cm}
\end{figure*}
Fig.~\ref{Fig:ResultsCondVsDiff}(a) shows the results for the conditional coder. Note that we only measure the rate required for the residual information and not for the motion field, since we want to compare the inter coding methods. Here, we see the large impact of the bottleneck. Since the curve with $\predChannels=64$ channels in the auxiliary latent space is much below the other curves, this curve is not shown completely in order to draw the focus on the other curves. For a bottleneck this tight, the conditional coder clearly performs worse than the residual coder. When we increase the number of channels and thereby widen the bottleneck, the coder performance increases and soon outperforms the residual coder. This increase of channels, however, comes with a large increase of model parameters, which has a negative impact on the complexity of the network.

\begin{figure}
	\centering
	\begin{tabular}{r}
		\scalebox{0.85}{
			\includegraphics{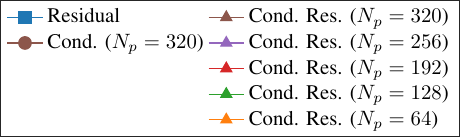}
		}\\
		\includegraphics{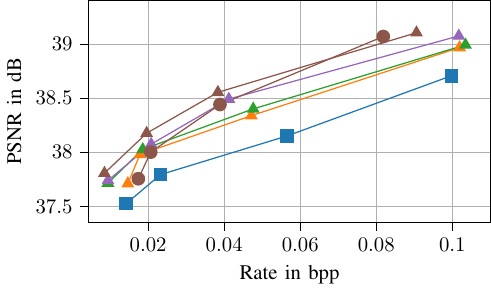}
	\end{tabular}\vspace{-0.3cm}
	\caption{Rate-distortion curve of the conditional residual coder. For comparison, also the residual coder and the conditional coder with a very wide bottleneck ($\predChannels=320$) are shown.\label{Fig:ResCodecNetResults}}
	\vspace{-0.5cm}
\end{figure}

From the theoretical derivations we deduced that the problems caused by the bottleneck become more severe when the prediction frame is of very high quality. To confirm that, we first remove all frames with a large prediction PSNR exceeding 35\dB from the test set. The results of this test are given in Fig.~\ref{Fig:ResultsCondVsDiff}(b).  Due to the worse prediction quality on average, the reconstruction quality drops, but compared to the performance of the residual coder, the conditional coders perform better. We see that indeed the bottlenecks are less severe. For $\predChannels=128$, the conditional coder already clearly outperforms the residual coder. Increasing $\predChannels$ increases the performance, however, after $\predChannels=192$ a saturation appears and a further increase of $\predChannels$ shows only very small effects. Apparently, with $\predChannels=192$, the prediction path is ``wide'' enough to pass on most signals with $\PSNR(\curV,\predV)\leq35\dB$ and further widening is not necessary. This results supports the notion ``information bottleneck'' for this effect.

As shown in Fig.~\ref{Fig:ResCodecNetResults}, for the conditional residual coder the picture looks completely different. Even when we test on the entire set, all coders perform better than the residual coder. We still see an increasing performance with wider bottlenecks, but the difference is much less severe than for the conditional coder. Here we have to also keep in mind that the number of parameters increases with $\predChannels$, which just by itself can also lead to an improved performance. 

\begin{figure*}
	\centering
	\begin{tabular}{cc}
		
		\scalebox{0.85}{
			\includegraphics{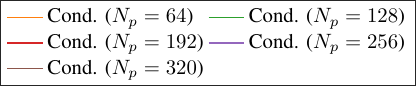}
		}
		&
		\scalebox{0.85}{
			\includegraphics{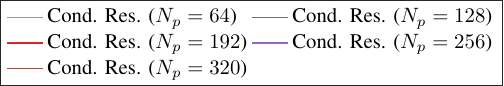}
		}
		\\
		\includegraphics{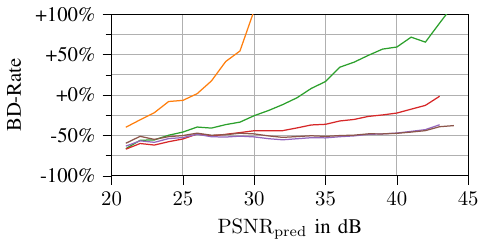}
		&
		\includegraphics{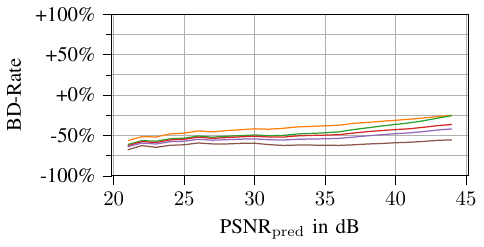}
		\vspace{-0.1cm}
		\\
		(a) Conditional coder & (b) Conditional residual coder
	\end{tabular}
	\caption{Bj\o ntegaard delta rate savings over the residual coder at different prediction qualities when varying the number $\predChannels$ of  channels in $f_p$.\label{Fig:BDRsCondCoderFilter}}
	\vspace{-0.5cm}
\end{figure*}

In order to show that the potential loss of the conditional coder is due to bottlenecks, we need to do a more refined analysis. Our model based evaluation predicts that bottlenecks are more severe for very good prediction signals. This was also confirmed in Fig.~\ref{Fig:ResultsCondVsDiff}(b). We therefore examine the performance at different prediction qualities, which we measure by the prediction PSNR 
\begin{equation}
\predPSNR\!=\!\mathrm{PSNR}\left(\curV,\predV\right).
\end{equation}
For each $\predPSNR$, we select sequences within a 6\dB interval of $\predPSNR$ around this value. For these sequences, we compute the Bj\o ntegaard delta rate~\cite{Bjontegaard2001_CalculationaveragePSNR,HerglotzOM2023_BjoentegaardBibleWhy} for each conditional coder relative to the residual coder. Fig.~\ref{Fig:BDRsCondCoderFilter} shows the results of this experiment. At the lowest point of $\predPSNR=21\dB$, which is a very poor prediction quality, the conditional coders, achieve large gains of more than 50\% rate savings for all bottlenecks, except for $\predChannels=64$. However, even for this very tight bottleneck, the gains are still large. Just with a slightly larger prediction PSNR, however, the coder with the second tightest bottleneck $\predChannels=128$ quickly gets a worse performance. At around 33\dBC, the residual coder outperforms this conditional coder. At this point the coder with $\predChannels=192$ still saves around 35\% rate. However, we see that also the performance for $\predChannels=192$ drops quickly after this point. With increasing prediction quality the conditional coding gain gets lower, while the drops are more severe for coders with tight bottlenecks. This again supports our theoretical considerations and findings that bottlenecks are more severe for frames with very good prediction signal.


Note that the coders perform all very similar up to a certain prediction PSNR, where the curve with the tightest bottleneck deviates from the others, rising upwards. For $\predChannels=128$, this occurs at around 25\dBC, and for $\predChannels=192$ at 30\dBC. This sudden drop of performance supports the interpretation that at a certain point the conditional path can no longer pass on all required information. This again validates the term ``information bottleneck'' and supports our model assumption in~(\ref{Eq:BNModel}).

Fig~\ref{Fig:BDRsCondCoderFilter}(b) shows the results for the conditional residual coder. Here, in contrast to the curves in Fig.~\ref{Fig:BDRsCondCoderFilter}(a), no bottleneck effects occur. The coders with wider bottlenecks show a better performance but unlike for the conditional coder, the conditional residual coder does not suddenly drop in performance when the prediction PSNR is too large. Also, as for the conditional coder, the performance decreases with increasing prediction PSNR. However, this is much less severe than for the conditional coder, and all coders achieve gains throughout all prediction PSNRs, even for a tight bottleneck. The slightly decreasing performance of the coders for increasing prediction PSNR is in line with the theory, since we showed that for the limiting case of perfect prediction, conditional, conditional residual, and residual coder all perform equally.

The experiments demonstrate that conditional residual coding provides a solution to the problem of bottlenecks which occur in conditional coding. We clearly showed that even tight bottlenecks do not harm the conditional coding gains much, which is in strong contrast to pure conditional coding, where a bottleneck of the same tightness severely worsened the results. 

\vspace{-0.1cm}
\subsection{Bottlenecks in the Decoder}
\begin{figure*}
		\begin{tabular}{cc}
			
			\scalebox{0.85}{
				\includegraphics{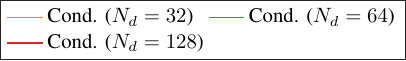}
			}
			&
			\scalebox{0.85}{
				\includegraphics{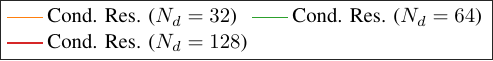}
			}
			\\
			\includegraphics{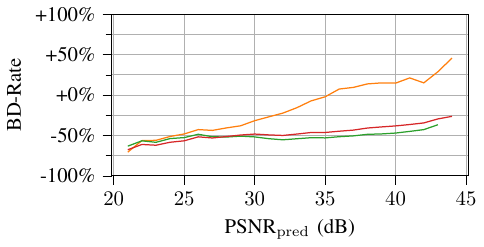}
			&
			\includegraphics{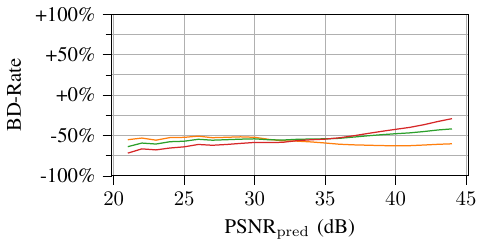}
			\vspace{-0.1cm}
			\\
			(a) Conditional coder & (b) Conditional residual coder
		\end{tabular}
		\caption{Bj\o ntegaard delta rate savings over the residual coder at different prediction qualities when varying the number $\decChannels$ of  channels in $f_d$ \label{Fig:VarDec}}
		\vspace{-0.5cm}
\end{figure*}

In the following, we want to demonstrate that bottlenecks can also occur in the decoder. To show that, we set $\predChannels=256$, which is a point where (as shown above) no major bottleneck effects occur. Instead, we vary the number of channels in the decoder $\decChannels$. We show the BD-rate over the prediction quality in Fig.~\ref{Fig:VarDec}. Note that we measure the BD-rate over the residual coder with 64 channels in the decoder, as before.

Again, we can observe a clear bottleneck in the conditional coder. For $\decChannels = 32$, the conditional coder still performs well for low prediction quality. At around 25\,dB, the rate savings suddenly decrease and for more than 35\,dB prediction PSNR, the conditional coder is outperformed by the residual coder. This is the same effect as in the previous experiment. The decoder does not have sufficient capacity to keep all relevant information from $\predV$. Therefore, from a certain prediction quality onwards, the coding efficiency drops. Fig.~\ref{Fig:VarDec}(b) shows that this effect does not occur in the conditional residual coder. This shows that while bottlenecks can also occur in the decoder of a conditional coder, the conditional residual coder again does not exhibit any bottleneck behavior. Even a coder with a relatively small $\decChannels=32$ can therefore be considered bottleneck-free. Our experiments therefore line up well with the theoretical results and prove their validity.

\vspace{-0.1cm}
\subsection{Complexity}
\begin{table}
	\centering
	\caption{Complexity for the different coders. We give the complexity required for encoding and for decoding, both in in kMAC/pixel and relative to the conditional coder with $\predChannels=64$. Note that conditional and conditional residual coder have the same complexity in terms of kMAC/pixel.\label{Tab:ComplexityMAC}}
	\begin{tabular}{ll|cc|cc}
		\toprule
		&& \multicolumn{2}{c|}{Encoder} & \multicolumn{2}{c}{Decoder} \\\midrule
		Residual&            & 49.0 &       79\%       & 39.3  &       78\%       \\\midrule
		\multirow{5}{1.3cm}{Conditional/ Cond. Res.}&$\predChannels=64$  & 62.4 &       100\%       & 50.5 &       100\%       \\
		&$\predChannels=128$ & 89.4 &       143\%       & 78.5 &       155\%       \\
		&$\predChannels=192$ & 134.2 &       215\%       & 123.3 &       244\%       \\
		&$\predChannels=256$ & 195.8 &       314\%       & 184.9 &       366\%       \\
		&$\predChannels=320$ & 274.2 &       439\%       & 263.3 &      521\% \\
		\bottomrule
	\end{tabular}
\vspace{-0.5cm}
\end{table}
When comparing different coding concepts, we also need to look at the complexity. In Table~\ref{Tab:ComplexityMAC}, we show the number of multiply accumulate (MAC) operations per pixel. This number is an estimate for the runtime of the coder. We show the results for the encoder (i.e., including all networks) and the decoder, including only the components needed for decoding. In this case, they are the decoder network of the entropy model, the decoder network and also the auxiliary encoder $\funcPredEnc$. Note that the complexity of conditional coder and conditional residual coder are equal, since they only differ in one subtraction and one addition per pixel.

When increasing $\predChannels$, the complexity also increases. For the conditional residual coder, we obtain a good performance also with $\predChannels=64$. We observe an increase of encoder complexity of 366\% when choosing $\predChannels=256$, which was the lowest $\predChannels$ for which we did not observe strong bottleneck effects.


Furthermore, looking at a recent video coder, such as FVC~\cite{HuLX2021_FVCNewFramework}, we note that the number of channels in the encoder and in the latent space is with 128 significantly lower than for image coders. The recent image coder ELIC~\cite{HeYP2022_ELICEfficientLearned} requires 320 channels in the latent space. Keep in mind that the conditional coder is essentially an image coder, i.e., the decoder requires the capacity to represent an image in each layer after the insertion of the prediction information. The conditional residual coder on the other hand is in essence still a residual coder, being able to operate with less channels. The conditional residual coder is therefore not only superior to a conditional coder in compression performance but also in complexity.

\section{Relationship to State-of-The-Art-Video Coders}
\label{Sec:RelToSoTA}
In the conditional coders proposed in the literature, we see different strategies, how bottlenecks were avoided, however, the problem has never been explicitly addressed. One example is the coder proposed by Ladune~\etal~\cite{LadunePH2020_ModeNetModeSelection}. As mentioned before, this coder also included ModeNet next to the conditional coder. ModeNet produces and transmits a mask with which parts of the image can be replaced by the prediction signal, similar to skip modes in traditional coders. Since typically only areas with very high prediction PSNR are skipped, this acts exactly on the areas where the bottleneck effects are strongest. This however comes at the cost of having to carry an additional coder for the mask, something which was subsequently alleviated~\cite{LadunePH2020_OpticalFlowMode}, where ModeNet was fused with the motion field estimation and transmission.

In DCVC~\cite{LiLL2021_DeepContextualVideo,LiLL2022_HybridSpatialTemporal}, the authors use a conditional coder, where the conditional information is added to the decoder at pixel resolution. This helps to alleviate the bottleneck, as there is no intermediate downsampling in the prediction path as in~\cite{LadunePH2020_ModeNetModeSelection}, but requires an extensive reconstruction network afterwards at pixel resolution to exploit the temporal redundancies at this level.

In~\cite{HoCC2022_CANFVCConditional}, the authors use augmented normalizing flows (ANFs) for conditional video compression. When analyzing the decoding process of their method, bottlenecks can also occur. However, their structure can be interpreted as a conditional residual coder, since the predicted signal is also used directly for reconstruction. The ANF-based approach is thereby a first step towards reducing bottlenecks.

These examples show that conditional coders in the literature are designed in a way that bottlenecks are alleviated. In this paper we provided a thorough theoretical foundation for such design choices. With the understanding of the underlying concepts that make these strategies necessary, further optimization of the coders can be more targeted to the bottleneck problems, thus enabling a more systematic design process. We furthermore suggest to include an evaluation for different prediction PSNRs, similar to Figs.~\ref{Fig:BDRsCondCoderFilter} and~\ref{Fig:ResCodecNetResults}(b) in the development, since they can be an indicator to potential bottleneck issues of a coder.

\section{Conclusion}
In this paper, we propose the concept of conditional residual inter coding. We draw the motivation for this approach from the analysis of conditional inter coding, which is prone to information bottlenecks in the prediction path. By modeling and analyzing these bottlenecks, we find that the existence of bottlenecks can cause the performance of a conditional coder to drop below the performance of a standard residual coder. By showing that a conditional residual coder is equivalent to the conditional coder in the ideal, bottleneck-free case, we elegantly solve the bottleneck problem since the conditional residual coder is guaranteed to perform better than the residual coder. In a subsequent information theoretical analysis, we show that the conditional residual coder outperforms the conditional coder in the presence of bottlenecks. We thereby present a solution which is proven to be better than both the standard residual coder and the currently popular conditional coder.

We furthermore present a series of experiments to support our theoretical findings. In a simple coding setup, we present a method to expose bottlenecks by analyzing the coder at different prediction qualities. For the conditional coder, we see the expected performance drop for high prediction qualities while the conditional residual coder maintains the high performance throughout all cases. We thereby detected information bottlenecks in the prediction path also experimentally and show that the conditional residual coder is free of these effects. This offers the possibility to design coders with much tighter bottlenecks, thereby presenting an opportunity to reduce the coder complexity. It becomes evident that the conditional residual coder does not have disadvantages over the conditional coder but can solve one of the main problems within this paradigm.

The concepts and results from this work can be used in future work to optimize video coding systems. To that end, the combination of conditional residual coding with other coding tools, e.g ModeNet~\cite{LadunePH2020_ModeNetModeSelection}, skip modes~\cite{ShiGW2022_AlphaVCHighPerformance}, or enhanced motion estimation methods~\cite{HuLG2022_CoarseFineDeep} should be investigated, as well as the behavior regarding error propagation through temporal prediction. Since the theoretical findings are unequivocally stating that conditional residual coding is the optimal concept for inter coding, gains can also be expected in the latest compression methods. The knowledge of bottleneck effects and theoretically sound mitigation strategies can contribute to more systematic and targeted improvements of conditional inter coding.

\appendix
\subsection{Theoretical Results for Lossy Conditional Coding}
In the following we show that $\RateCond(\dist) \leq \RateRes(\dist)$. To compare the two rate-distortion functions, we start with Bayes' law:
\begin{equation}
\begin{split}
H(\resD,\recresD) &+ H(\curD,\recD,\predD|\resD,\recresD) \\=& H(\curD,\recD,\predD) + \underbrace{H(\resD,\recresD|\curD,\recD,\predD)}_{=0}
\end{split}
\end{equation}
The last term is equal to zero because together with the prediction signal $\predV$, $\curV$ and $\recV$ completely determine $\resV$ and $\recresV$.
We can further rearrange the equation to obtain
\begin{subequations}
	\begin{align}
	H(\resD,\recresD)=&H(\curD,\recD,\predD) - H(\curD,\recD,\predD|\resD,\recresD)\\
	=&H(\curD,\recD|\predD) + H(\predD) \nonumber \\ &-\underbrace{H(\curD,\recD|\resD,\recresD,\predD)}_{=0}-\underbrace{H(\predD|\resD,\recresD)}_{=H(\predD|\resD)}\\
	=&H(\curD,\recD|\predD) + I(\predD;\resD)
	\end{align}
\end{subequations}

With this equation, we can now connect $H(\resD,\recresD)$ and
$H(\curD,\recD|\predD)$. We continue with:
\begin{equation}
\begin{split}
H(\resD,\recresD) &= I(\resD;\recresD) + H(\resD|\recresD) + \underbrace{H(\recresD|\resD)}_{=0} \\
&=I(\curD;\recD|\predD) + H(\curD|\recD,\predD) \\&~~~+ \underbrace{H(\recD|\curD,\predD)}_{=0} + I(\predD;\resD)
\end{split}
\end{equation}
We can therefore write
\begin{subequations}
	\begin{align}
	I(\resD;\recresD)=&I(\curD;\recD|\predD)+I(\predD;\resD)\nonumber\\&+H(\curD|\recD,\predD)-H(\resD|\recresD)\\
	=&I(\curD;\recD|\predD) + I(\predD;\resD) \nonumber \\&+ H(\resD|\recD,\predD,\recresD) - H(\resD|\recresD)\label{Eq:FinalRDCurveA}\\
	=&I(\curD;\recD|\predD) + I(\predD;\resD) - I(\resD;\recD,\predD|\recresD)\\
	=&I(\curD;\recD|\predD) + I(\predD;\resD) - I(\resD;\predD|\recresD)\label{Eq:FinalRDCurve}
	\end{align}
\end{subequations}
In (\ref{Eq:FinalRDCurveA}), we exploit the identity $H(\curD|\recD,\predD)=H(\resD|\recD,\predD)=H(\resD|\recD,\predD,\recresD)$, which follows from the fact that $\curV$ and $\resV$ carry the same information when $\predV$ is known and that $\recresV$ is known when $\recV$ and $\predV$ are known. In (\ref{Eq:FinalRDCurve}), we exploit that when $\recresV$ is known, knowing $\predV$ also means knowing $\recV$. Therefore $\recV$ can not contribute to the mutual information. We can now plug in the result in the rate distortion function~(\ref{Eq:RDEqResidual}) and obtain:
\begin{subequations}
	\begin{align}
	\RateRes&(\dist) = \!\!\!\!\min_{p\in\PSet(\dist)}\big(I(\curD;\recD|\predD) + I(\predD;\resD) - I(\resD;\predD|\recresD)\big)\\
	\geq& I(\predD;\resD) + \!\!\!\!\min_{p\in\PSet(\dist)}I(\curD;\recD|\predD) +  \!\!\!\!\min_{p\in\PSet(\dist)} - I(\resD;\predD|\recresD)\label{Eq:PullOutMutualInformation}\\
	=&\RateCond(\dist)\!+\!I(\predD;\resD)\!-\!\!\!\max_{p\in\PSet(\dist)} I(\resD;\predD|\recresD)\\\geq&\RateCond(\dist)\label{Eq:FinalInequality}
	\end{align}
\end{subequations}
In (\ref{Eq:PullOutMutualInformation}), we use the fact that the marginal $p_\mathrm{m}(\curV,\predV)$ of all distributions in $\PSet(\dist)$ is constant for a given scenario. Since $I(\predD;\resD)$ is fully described by this marginal, we can exclude this term from the minimization operation. Since $I(\predD;\resD) \geq I(\resD;\predD|\recresD)$ holds in general, also $I(\predD;\resD) \geq \max_{p\in\PSet(\dist)} I(\resD;\predD|\recresD)$ holds, thus proving the inequality  in (\ref{Eq:FinalInequality}), which proves (\ref{Eq:MainLossyX}).

\bibliographystyle{IEEEbib}

\begin{thebibliography}{10}
	
	\bibitem{TodericiOH2016_VariableRateImage}
	G.~Toderici, S.~M. O'Malley, S.~J. Hwang, D.~Vincent, D.~Minnen, S.~Baluja,
	M.~Covell, and R.~Sukthankar,
	\newblock ``Variable rate image compression with recurrent neural networks,''
	\newblock in {\em Proc. International Conference on Learning Representations
		({ICLR})}, Y.~Bengio and Y.~LeCun, Eds., 2016.
	
	\bibitem{BalleLS2017_Endendoptimized}
	J.~Ball{\'e}, V.~Laparra, and E.~P. Simoncelli,
	\newblock ``End-to-end optimized image compression,''
	\newblock in {\em Proc. International Conference on Learning Representations
		(ICLR)}, Apr. 2017, pp. 1--27.
	
	\bibitem{BalleMS2018_Variationalimagecompression}
	J.~Ballé, D.~Minnen, S.~Singh, S.~J. Hwang, and N.~Johnston,
	\newblock ``Variational image compression with a scale hyperprior,''
	\newblock in {\em Proc. International Conference on Learning Representations
		({ICLR})}, 2018, pp. 1--47.
	
	\bibitem{LuOX2018_DVCEndend}
	G.~Lu, W.~Ouyang, D.~Xu, X.~Zhang, C.~Cai, and Z.~Gao,
	\newblock ``{DVC}: An end-to-end deep video compression framework,''
	\newblock in {\em Proc. {IEEE}/{CVF} Conference on Computer Vision and Pattern
		Recognition ({CVPR})}, June 2019, pp. 10998--11007.
	
	\bibitem{HuLX2021_FVCNewFramework}
	Z.~Hu, G.~Lu, and D.~Xu,
	\newblock ``{FVC}: A new framework towards deep video compression in feature
	space,''
	\newblock in {\em Proc. IEEE/CVF Conference on Computer Vision and Pattern
		Recognition (CVPR)}, June 2021, pp. 1502--1511.
	
	\bibitem{SullivanOH2012_OverviewHighEfficiency}
	G.~J. Sullivan, J.-R. Ohm, W.-J. Han, and T.~Wiegand,
	\newblock ``Overview of the high efficiency video coding ({HEVC}) standard,''
	\newblock {\em IEEE Transactions on Circuits and Systems for Video Technology},
	vol. 22, no. 12, pp. 1649--1668, Dec. 2012.
	
	\bibitem{BrossWY2021_OverviewVersatileVideo}
	B.~Bross, Y.-K. Wang, Y.~Ye, S.~Liu, J.~Chen, G.~J. Sullivan, and J.-R. Ohm,
	\newblock ``Overview of the versatile video coding ({VVC}) standard and its
	applications,''
	\newblock {\em {IEEE} Transactions on Circuits and Systems for Video
		Technology}, vol. 31, no. 10, pp. 3736--3764, Oct. 2021.
	
	\bibitem{HanLM2021_TechnicalOverviewAV1}
	J.~Han, B.~Li, D.~Mukherjee, C.-H. Chiang, A.~Grange, C.~Chen, H.~Su,
	S.~Parker, S.~Deng, U.~Joshi, Y.~Chen, Y.~Wang, P.~Wilkins, Y.~Xu, and
	J.~Bankoski,
	\newblock ``A technical overview of {AV}1,''
	\newblock {\em Proceedings of the {IEEE}}, vol. 109, no. 9, pp. 1--28, 2021.
	
	\bibitem{LadunePH2020_ModeNetModeSelection}
	T.~Ladune, P.~Philippe, W.~Hamidouche, L.~Zhang, and O.~Deforges,
	\newblock ``Modenet: Mode selection network for learned video coding,''
	\newblock in {\em Proc. {IEEE} International Workshop on Machine Learning for
		Signal Processing ({MLSP})}, Sept. 2020, pp. 1--6.
	
	\bibitem{LiLL2021_DeepContextualVideo}
	J.~Li, B.~Li, and Y.~Lu,
	\newblock ``Deep contextual video compression,''
	\newblock in {\em Proc. Advances in Neural Information Processing Systems
		(NeurIPS)}, M.~Ranzato, A.~Beygelzimer, Y.~Dauphin, P.~Liang, and J.~W.
	Vaughan, Eds. 2021, vol.~34, pp. 18114--18125, Curran Associates, Inc.
	
	\bibitem{ShengLL2022_TemporalContextMining}
	X.~Sheng, J.~Li, B.~Li, L.~Li, D.~Liu, and Y.~Lu,
	\newblock ``Temporal context mining for learned video compression,''
	\newblock {\em {IEEE} Transactions on Multimedia}, pp. 1--12, 2022.
	
	\bibitem{LiLL2022_HybridSpatialTemporal}
	J.~Li, B.~Li, and Y.~Lu,
	\newblock ``Hybrid spatial-temporal entropy modelling for neural video
	compression,''
	\newblock in {\em Proc. {ACM} International Conference on Multimedia}, Oct.
	2022, pp. 1503--1511.
	
	\bibitem{LiLL2023_NeuralVideoCompression}
	J.~Li, B.~Li, and Y.~Lu,
	\newblock ``Neural video compression with diverse contexts,''
	\newblock in {\em Proc. IEEE/CVF Conference on Computer Vision and Pattern
		Recognition (CVPR)}, June 2023, pp. 22616--22626.
	
	\bibitem{MentzerTM2022_VCTVideoCompression}
	F.~Mentzer, G.~D. Toderici, D.~Minnen, S.~Caelles, S.~J. Hwang, M.~Lucic, and
	E.~Agustsson,
	\newblock ``{VCT}: A video compression transformer,''
	\newblock in {\em Proc. Advances in Neural Information Processing Systems
		(NeurIPS)}, S.~Koyejo, S.~Mohamed, A.~Agarwal, D.~Belgrave, K.~Cho, and
	A.~Oh, Eds. Dec. 2022, vol.~35, pp. 13091--13103, Curran Associates, Inc.
	
	\bibitem{SohnLY2015_LearningStructuredOutput}
	K.~Sohn, H.~Lee, and X.~Yan,
	\newblock ``Learning structured output representation using deep conditional
	generative models,''
	\newblock in {\em Proc. Advances in Neural Information Processing Systems
		(NeurIPS)}, pp. 3483--3491. 2015.
	
	\bibitem{HosoeYK2018_OfflineTextIndependent}
	M.~{Hosoe}, T.~{Yamada}, K.~{Kato}, and K.~{Yamamoto},
	\newblock ``Offline text-independent writer identification based on
	writer-independent model using conditional autoencoder,''
	\newblock in {\em Proc. International Conference on Frontiers in Handwriting
		Recognition ({ICFHR})}, Aug. 2018, pp. 441--446.
	
	\bibitem{BrandSK2020_IntraFrameCoding}
	F.~Brand, J.~Seiler, and A.~Kaup,
	\newblock ``Intra-frame coding using a conditional autoencoder,''
	\newblock {\em {IEEE} Journal of Selected Topics in Signal Processing}, vol.
	15, no. 2, pp. 354--365, Feb. 2021.
	
	\bibitem{BrandSK2022_BenefitsChallengesConditional}
	F.~Brand, J.~Seiler, and A.~Kaup,
	\newblock ``On benefits and challenges of conditional interframe video coding
	in light of information theory,''
	\newblock in {\em Proc. Picture Coding Symposium ({PCS})}, Dec. 2022, pp.
	289--293.
	
	\bibitem{SlepianW1973_Noiselesscodingcorrelated}
	D.~Slepian and J.~Wolf,
	\newblock ``Noiseless coding of correlated information sources,''
	\newblock {\em IEEE Transactions on Information Theory}, vol. 19, no. 4, pp.
	471--480, 1973.
	
	\bibitem{AgustssonMJ2020_ScaleSpaceFlow}
	E.~Agustsson, D.~Minnen, N.~Johnston, J.~Balle, S.~J. Hwang, and G.~Toderici,
	\newblock ``Scale-space flow for end-to-end optimized video compression,''
	\newblock in {\em Proc. IEEE/CVF Conference on Computer Vision and Pattern
		Recognition (CVPR)}, June 2020, pp. 8503--8512.
	
	\bibitem{HuLG2022_CoarseFineDeep}
	Z.~Hu, G.~Lu, J.~Guo, S.~Liu, W.~Jiang, and D.~Xu,
	\newblock ``Coarse-to-fine deep video coding with hyperprior-guided mode
	prediction,''
	\newblock in {\em Proc. IEEE/CVF Conference on Computer Vision and Pattern
		Recognition (CVPR)}, June 2022, pp. 5921--5930.
	
	\bibitem{LinJZ2023_DMVCDecomposedMotion}
	K.~Lin, C.~Jia, X.~Zhang, S.~Wang, S.~Ma, and W.~Gao,
	\newblock ``{DMVC}: Decomposed motion modeling for learned video compression,''
	\newblock {\em {IEEE} Transactions on Circuits and Systems for Video
		Technology}, vol. 33, no. 7, pp. 3502--3515, July 2023.
	
	\bibitem{YangTG2023_AdvancingLearnedVideo}
	R.~Yang, R.~Timofte, and L.~V. Gool,
	\newblock ``Advancing learned video compression with in-loop frame
	prediction,''
	\newblock {\em {IEEE} Transactions on Circuits and Systems for Video
		Technology}, vol. 33, no. 5, pp. 2410--2423, May 2023.
	
	\bibitem{WiegandSB2003_OverviewH.264/AVCvideo}
	T.~Wiegand, G.~J. Sullivan, G.~Bj{\o}ntegaard, and A.~Luthra,
	\newblock ``Overview of the {H}.264/{AVC} video coding standard,''
	\newblock {\em IEEE Transactions on Circuits and Systems for Video Technology},
	vol. 13, no. 7, pp. 560--576, July 2003.
	
	\bibitem{MukherjeeBG2013_latestopensource}
	D.~Mukherjee, J.~Bankoski, A.~Grange, J.~Han, J.~Koleszar, P.~Wilkins, Y.~Xu,
	and R.~Bultje,
	\newblock ``The latest open-source video codec {VP}9 - an overview and
	preliminary results,''
	\newblock in {\em Proc. Picture Coding Symposium ({PCS})}, Dec. 2013, pp.
	390--393.
	
	\bibitem{LadunePH2020_OpticalFlowMode}
	T.~Ladune, P.~Philippe, W.~Hamidouche, L.~Zhang, and O.~Deforges,
	\newblock ``Optical flow and mode selection for learning-based video coding,''
	\newblock in {\em Proc. {IEEE} Workshop on Multimedia Signal Processing}, Sept.
	2020, pp. 1--6.
	
	\bibitem{LaduneP2022_AIVCArtificialIntelligence}
	T.~Ladune and P.~Philippe,
	\newblock ``Aivc: Artificial intelligence based video codec,''
	\newblock in {\em Proc. {IEEE} International Conference on Image Processing
		(ICIP)}, Oct. 2022, pp. 316--320.
	
	\bibitem{BrandSK2022_PFrameCoding}
	F.~Brand, J.~Seiler, and A.~Kaup,
	\newblock ``P-frame coding with generalized difference: A novel conditional
	coding approach,''
	\newblock in {\em Proc. IEEE International Conference on Image Processing
		(ICIP)}, Oct. 2022, pp. 1266--1270.
	
	\bibitem{ChenWL2023_CompactTemporalTrajectory}
	B.~Chen, Z.~Wang, B.~Li, S.~Wang, and Y.~Ye,
	\newblock ``Compact temporal trajectory representation for talking face video
	compression,''
	\newblock {\em {IEEE} Transactions on Circuits and Systems for Video
		Technology}, pp. 1--1, 2023.
	
	\bibitem{HoCC2022_CANFVCConditional}
	Y.-H. Ho, C.-P. Chang, P.-Y. Chen, A.~Gnutti, and W.-H. Peng,
	\newblock ``{CANF}-{VC}: Conditional augmented normalizing flows for~video
	compression,''
	\newblock in {\em Proc. European Conference on Computer Vision (ECCV)}, 2022,
	pp. 207--223.
	
	\bibitem{ViterbiO1979_PrinciplesDigitalCommunication}
	A.~J. Viterbi and J.~K. Omura,
	\newblock {\em Principles of Digital Communication and Coding},
	\newblock McGraw-Hill, 1979.
	
	\bibitem{LiLD2020_SpatialChannelContext}
	C.~Li, J.~Luo, W.~Dai, C.~Li, J.~Zou, and H.~Xiong,
	\newblock ``Spatial-channel context-based entropy modeling for end-to-end
	optimized image compression,''
	\newblock in {\em Proc. International Conference on Visual Communications and
		Image Processing (VCIP)}, 2020, pp. 222--225.
	
	\bibitem{Mentzer_clic2020devkit}
	F.~Mentzer,
	\newblock ``{clic2020-devkit},''
	\url{https://github.com/fab-jul/clic2020-devkit}, 2020,
	\newblock {Last accessed}: 15.11.2021 17:18.
	
	\bibitem{KingmaB2015_AdamMethodStochastic}
	D.~P. Kingma and J.~Ba,
	\newblock ``Adam: A method for stochastic optimization,''
	\newblock in {\em Proc. International Conference on Learning Representations
		(ICLR)}, May 2015, pp. 1--15.
	
	\bibitem{MinnenBT2018_JointAutoregressiveHierarchical}
	D.~Minnen, J.~Ball\'{e}, and G.~D. Toderici,
	\newblock ``Joint autoregressive and hierarchical priors for learned image
	compression,''
	\newblock in {\em Proc. Advances in Neural Information Processing Systems
		(NeurIPS)}, Dec. 2018, vol.~31, pp. 1--10.
	
	\bibitem{BegaintRF2020_CompressAIPyTorchlibrary}
	J.~B{\'e}gaint, F.~Racap{\'e}, S.~Feltman, and A.~Pushparaja,
	\newblock ``Compressai: a pytorch library and evaluation platform for
	end-to-end compression research,''
	\newblock {\em arXiv preprint arXiv:2011.03029}, pp. 1--19, Nov. 2020.
	
	\bibitem{Bjontegaard2001_CalculationaveragePSNR}
	G.~Bj{\o}ntegaard,
	\newblock ``Calculation of average {PSNR} differences between {RD}-curves,
	{VCEG-M}33,''
	\newblock {\em 13th Meeting of the Video Coding Experts Group ({VCEG})}, pp.
	1--5, Jan. 2001.
	
	\bibitem{HerglotzOM2023_BjoentegaardBibleWhy}
	C.~Herglotz, H.~Och, A.~Meyer, G.~Ramasubbu, L.~Eichermüller, M.~Kränzler,
	F.~Brand, K.~Fischer, D.~T. Nguyen, A.~Regensky, and A.~Kaup,
	\newblock ``The bjøntegaard bible -- why your way of comparing video codecs
	may be wrong,''
	\newblock {\em Arxiv Preprint, arxiv:2304.12852v1 (Accepted for Publication in
		IEEE Transactions on Image Processing)}, 2023.
	
	\bibitem{HeYP2022_ELICEfficientLearned}
	D.~He, Z.~Yang, W.~Peng, R.~Ma, H.~Qin, and Y.~Wang,
	\newblock ``Elic: Efficient learned image compression with unevenly grouped
	space-channel contextual adaptive coding,''
	\newblock in {\em Proc. Conference on Computer Vision and Pattern Recognition
		(CVPR)}, June 2022, pp. 5718--5727.
	
	\bibitem{ShiGW2022_AlphaVCHighPerformance}
	Y.~Shi, Y.~Ge, J.~Wang, and J.~Mao,
	\newblock ``Alphavc: High-performance and efficient learned video
	compression,''
	\newblock in {\em Proc. European Conference on Computer Vision ({ECCV})},
	S.~Avidan, G.~Brostow, M.~Ciss{\'e}, G.~M. Farinella, and T.~Hassner, Eds.,
	Cham, Oct. 2022, pp. 616--631, Springer Nature Switzerland.
	
\end{thebibliography}

\begin{IEEEbiography}[{\includegraphics[width=1in,height=1.25in,clip,keepaspectratio]{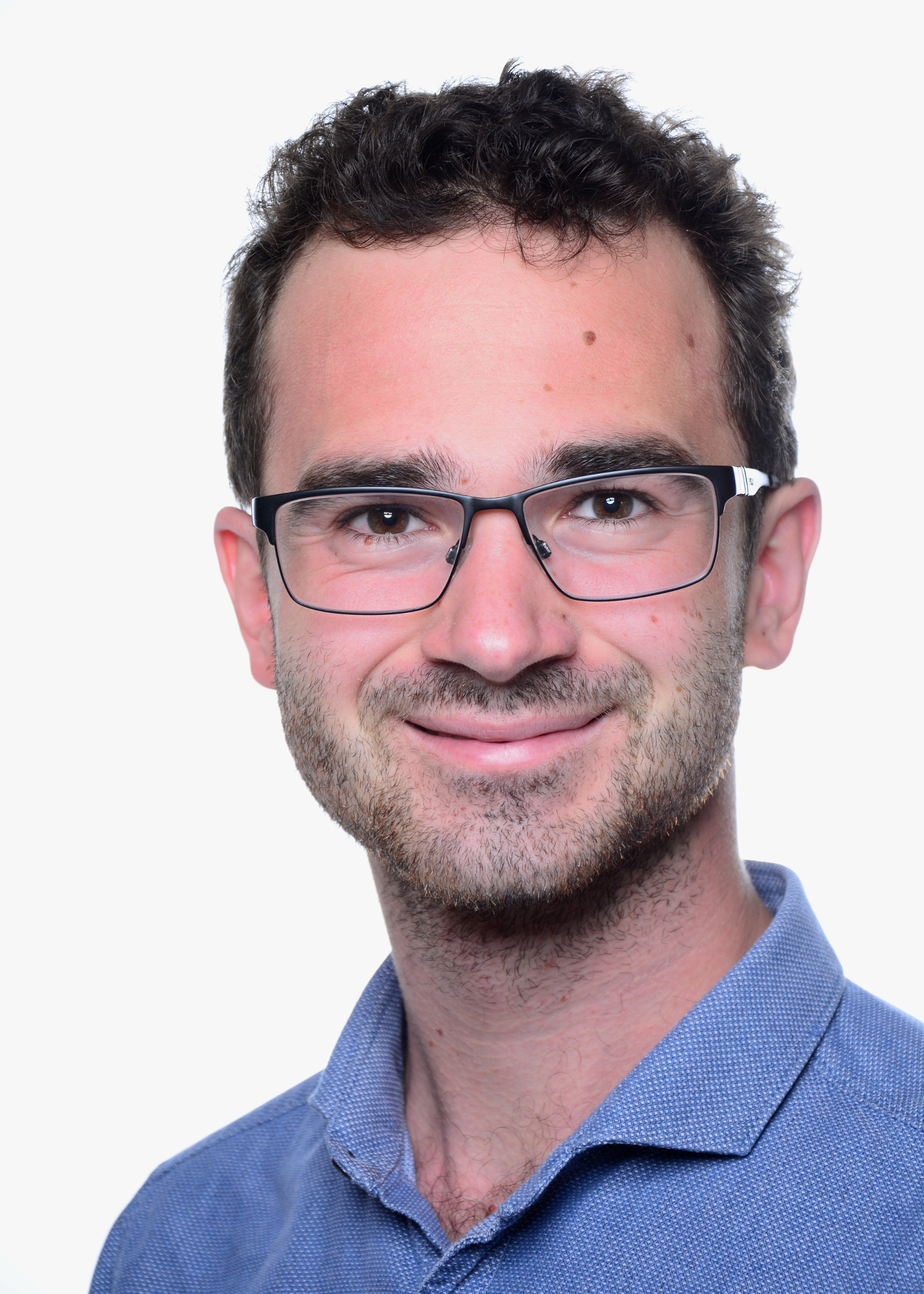}}]{Fabian Brand}
	Fabian Brand (Graduate Student Member, IEEE) received the master’s degree in electrical engineering from Friedrich-Alexander-Universität Erlangen-Nürnberg (FAU), Germany, in 2018. During his bachelor’s, he worked on methods for frame-rate-conversion of video sequences, and during his master’s, he researched automated harmonic analysis of classical music and style classification. Since 2019, he has been a Researcher with the Chair of Multimedia Communications and Signal Processing, FAU, where he conducts research on methods for video compression and deep learning. For his work, among others, he received the Best Paper Award of the Picture Coding Symposium (PCS) 2019.\end{IEEEbiography}

\begin{IEEEbiography}[{\includegraphics[width=1in,height=1.25in,clip,keepaspectratio]{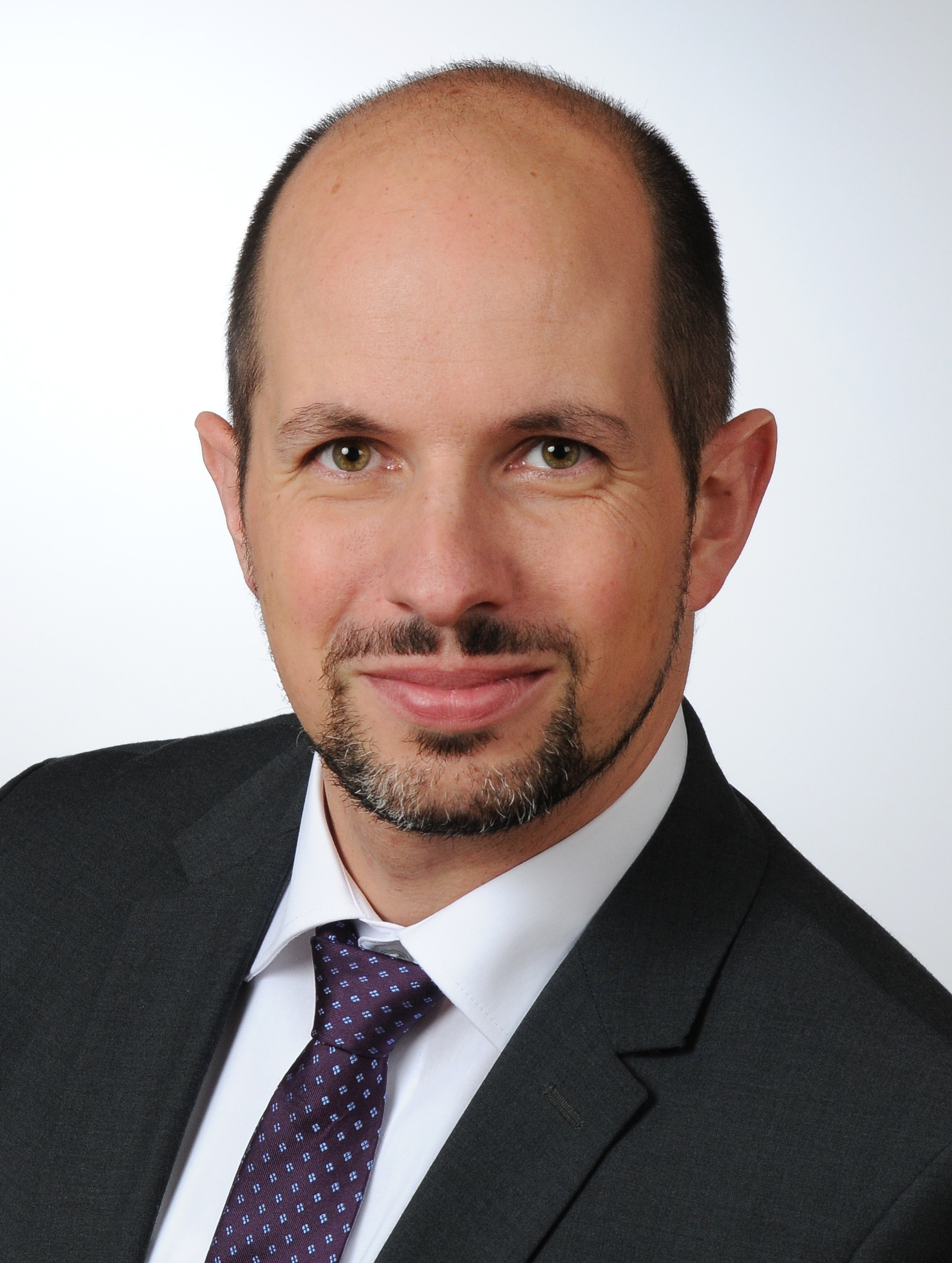}}]{J\"{u}rgen Seiler}
	J{\"u}rgen Seiler (Senior Member IEEE) is senior scientist and lecturer at the Chair of Multimedia Communications and Signal Processing at the Friedrich-Alexander Universit{\"a}t Erlangen-N{\"u}rnberg, Germany. There, he also received his habilitation degree in 2018, his doctoral degree in 2011, and his diploma degree in Electrical Engineering, Electronics and Information Technology in 2006.
	
	He received the dissertation award of the Information Technology Society of the German Electrical Engineering Association as well as the dissertation award of the Staedtler-Foundation, both in 2012. In 2007, he received diploma awards from the Institute of Electrical Engineering, Electronics and Information Technology, Erlangen, as well as from the German Electrical Engineering Association. He also received scholarships from the German National Academic Foundation and the Lucent Technologies Foundation. He is the co-recipient of five best paper awards and he has authored or co-authored more than 120 technical publications.
	
	His research interests include image and video signal processing, signal reconstruction and coding, signal transforms, and linear systems theory.\end{IEEEbiography}

\begin{IEEEbiography}[{\includegraphics[width=1in,height=1.25in,clip,keepaspectratio]{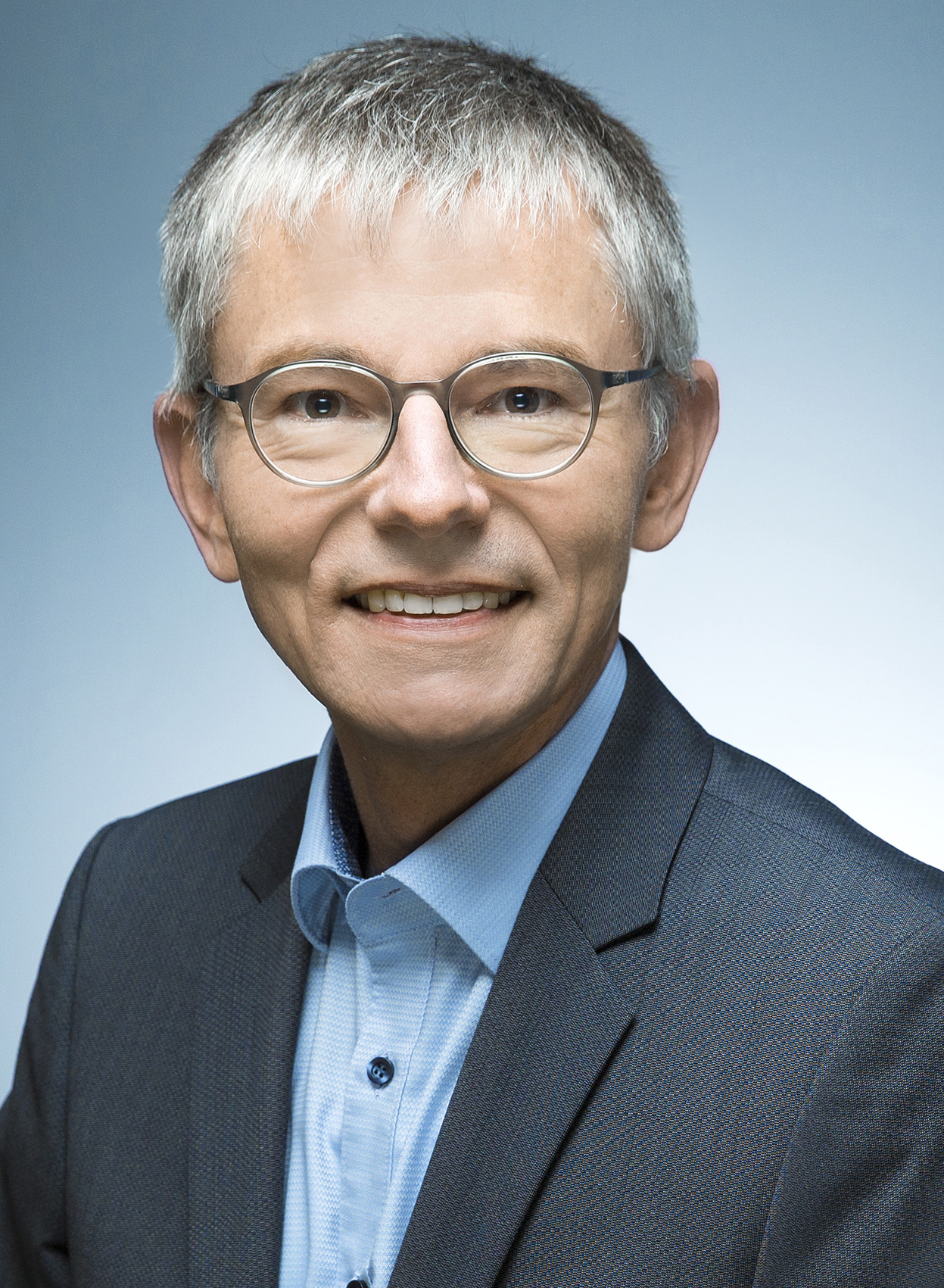}}]{Andr\'{e} Kaup}
	André Kaup (Fellow, IEEE) received the Dipl.-Ing. and Dr.-Ing. degrees in electrical engineering from RWTH Aachen University, Aachen, Germany, in 1989 and 1995, respectively.
	
	He joined Siemens Corporate Technology, Munich, Germany, in 1995, and became the Head of the Mobile Applications and Services Group in 1999. Since 2001, he has been a Full Professor and the Head of the Chair of Multimedia Communications and Signal Processing at Friedrich-Alexander University Erlangen-Nürnberg (FAU), Germany. From 2005 to 2007 he was Vice Speaker of the DFG Collaborative Research Center 603. From 2015 to 2017, he served as the Head of the Department of Electrical Engineering and Vice Dean of the Faculty of Engineering at FAU. He has authored around 450 journal and conference papers and has over 120 patents granted or pending. His research interests include image and video signal processing and coding, and multimedia communication.
	
	Dr. Kaup is a member of the IEEE Image, Video, and Multidimensional Signal Processing Technical Committee and a member of the Scientific Advisory Board of the German VDE/ITG. He is an IEEE Fellow, a member of the Bavarian Academy of Sciences and Humanities, and a member of the European Academy of Sciences and Arts. He is a member of the Editorial Board of the IEEE Circuits and Systems Magazine. He was a Siemens Inventor of the Year 1998 and obtained the 1999 ITG Award. He received several IEEE best paper awards, including the Paul Dan Cristea Special Award in 2013, and his group won the Grand Video Compression Challenge from the Picture Coding Symposium 2013. The Faculty of Engineering with FAU and the State of Bavaria honored him with Teaching Awards, in 2015 and 2020, respectively. He served as an Associate Editor of the IEEE Transactions on Circuits and Systems for Video Technology. He was a Guest Editor of the IEEE Journal of Selected Topics in Signal Processing. \end{IEEEbiography}

\end{document}